\documentclass[review]{elsarticle}

\usepackage{lineno,hyperref}
\usepackage{tikz-cd}
\usepackage{dsfont}
\usepackage{amsfonts}
\usepackage{amsmath}
\usepackage{slashed}
\usepackage{graphics}
\usepackage{amsmath}
\usepackage{amssymb}
\usepackage{tikz-cd}
\usepackage[compat=1.1.0]{tikz-feynman}
\usepackage{slashed}
\usepackage{setspace}
\modulolinenumbers[5]

\makeatletter
\def\ps@pprintTitle{%
 \let\@oddhead\@empty
 \let\@evenhead\@empty
 \def\@oddfoot{\centerline{\thepage}}%
 \let\@evenfoot\@oddfoot}
\makeatother

\journal{EPJ-C - doi: 10.1140/epjc/s10052-019-7301-7}
\begin{document}

\begin{frontmatter}

\title{\textbf{Homogeneously Modified Special Relativity (HMSR)}\\ A new possible way to introduce an isotropic Lorentz Invariance Violation in particle Standard Model}

\tnotetext[t1]{Corresponding author email: marco.torri@unimi.it\\ \\ \\ EPJ-C 79 (2019) no.9, 808 - doi: 10.1140/epjc/s10052-019-7301-7}

\author[Unimi]{M.D.C. Torri*\corref{mycorrespondingauthor}}

\author[Unimi]{V. Antonelli}

\author[Unimi]{L. Miramonti}

\address[Unimi]{Dipartimento di Fisica\\
               Universit\'a degli Studi e INFN\\
               via Celoria 16, 20133, Milano, Italy}

\begin{abstract}
This work explores a Standard Model extension possibility, that violates Lorentz invariance, preserving the space-time isotropy and homogeneity. In this sense HMSR represents an attempt to introduce an isotropic Lorentz Invariance Violation in the elementary particle SM. The theory is constructed starting from a modified kinematics, that takes into account supposed quantum effects due to interaction with the space-time background. The space-time structure itself is modified, resulting in a pseudo-Finsler manifold. The SM extension here provided is inspired by the effective fields theories, but it preserves covariance, with respect to newly introduced modified Lorentz transformations. Geometry perturbations are not considered as universal, but particle species dependent. Non universal character of the amended Lorentz transformations allows to obtain visible physical effects, detectable in experiments by comparing different perturbations related to different interacting particles species.
\end{abstract}


\end{frontmatter}

\linenumbers
\section{Introduction}
Most of Lorentz Invariance Violating (LIV) theories are characterized by a modification of the free particle kinematics. This effect is supposed to be caused by the interaction of the free propagating particle with the space-time quantized background structure. In fact one expects that Planck-scale interactions could manifest themselves in a "low" energy scenario as tiny residual effects, that can modify standard physics. Several candidates theories, such as Standard Model Extension (SME) \cite{Koste2,Koste3,Koste4,Koste5,Kostel3}, Double Special Relativity (DSR) \cite{Amelino,Amelino2,Amelino3,Amelino4,AmelinoCamelia,AmelinoCamelia2,AmelinoCamelia3,AmelinoCameliafin4,Smolin1,Smolin2}, Very Special Relativity (VSR) \cite{Glash1,Glash2}, have been proposed. All of these theories share the feature of considering modified dispersion relations for free particles, with the amended form $E^2-(1-f(p))p^2=m^2$ and $f(p)=\sum_{k=1}\alpha_{k}(E_{P})p^k$. The resulting space-time geometry acquires an energy-momentum dependence, that can be described resorting to Finsler geometry \cite{Koste1,Liberati1,Edwards,Lammerzahl,Bubuianu,Schreck}. Some proposed scenarios main feature consists in providing background structures, that introduce preferred directions to violate Lorentz Invariance, as in SME \cite{Koste2}. This characteristic implies that the space-time is no more isotropic and therefore an inertial observers privileged class must exist. Quantum Gravity effects could emerge in a symmetry breaking fashion. In fact, spontaneous symmetry breaking is a useful concept, employed for example in particle physics. One, therefore, can suppose that, even in the high energy limit, the Planck scenario, quantum gravity could present the same mechanism, breaking the Lorentz symmetry. The introduction of a privileged reference frame might not be a real problem, even if it might result conceptually difficult. Some studies, for example, attempt to correlate this privileged reference frame with the natural one, used for the description of the Cosmic Microwave Background Radiation (CMBR). But there are apparently not physical reasons that can justify any connection between a supposed quantum phenomenon of the Planck scale, with the CMBR classical physics description. In fact, nowadays, it is not clear how to introduce and justify these preferred inertial observers.\\
In this work, to preserve the idea of space-time isotropy, a possible way to introduce a LIV theory, without a preferred class of inertial observers, is explored. The Lorentz symmetry is therefore only modified, as in DSR theories \cite{AmelinoCamelia, AmelinoCamelia2,AmelinoCamelia3}. Hence the idea of space-time isotropy results restored respect to the amended Lorentz transformations, like in DSR theories \cite{AmelinoCamelia}.  Lorentz symmetry perturbations are not introduced in an universal way, instead every particle species presents its personal modification, as suggested in \cite{Koste5}. This corresponds, for the high energy limit, to a redefinition of the maximum attainable velocity, different for every particle type, as for example in \cite{Glash1}. Moreover, in this way, it results possible to predict detectable physical effects, without the introduction of exotic mechanisms.

\section{Modified Dispersion Relations}
In Homogeneously Modified Special relativity (HMSR), to geometrize the supposed interaction of massive particles with the background, LIV is introduced, exploiting the possibility to perturb kinematics. This approach consists in modifying the dispersion relations describing the free particle propagation. The Dispersion Relations of standard physics, written using the Minkowski metric as $E^2-|\overrightarrow{p}|^2=m^2$ $\Rightarrow\;p_{\mu}\eta^{\mu\nu}p_{\nu}=m^2$, are modified, following \cite{Torri,Antonelli}, to a more general case of Modified Dispersion Relations (MDR):
\begin{equation}
\label{c1}
MDR(p)=E^2-\left(1-f\left(\frac{|\overrightarrow{p}|}{E}\right)-g\left(\frac{\overrightarrow{p}}{E}\right)\right)|\overrightarrow{p}|^2=m^2
\end{equation}
The $f$ perturbation function preserves the rotational invariance of the MDR, the $g$ one instead breaks this symmetry, introducing a preferred direction in space-time. It is even important to stress that the lack of distinction between particles and antiparticles in MDR means that one is dealing with a CPT even theory, in fact the dispersion relations do not present a dependence on particle helicity or spin. In order to contemplate a CPT odd model extension, it is therefore sufficient to include this kind of dependence in MDR formulation. Since the publication of the Greenberg paper \cite{Greenberg}, it is well known that LIV does not imply CPT violation. The opposite statement is supposed true in the same work \cite{Greenberg}, but this point is controversial. The idea that CPT violation automatically implies LIV is widely debated in literature \cite{Discussion-LIV-CPT,Discussion-LIV-CPT2,Discussion-LIV-CPT3,Discussion-LIV-CPT4} and it was confuted, for example, in \cite{Dolgov-et-al}.\\
In order to preserve the geometrical origin of MDR, the perturbation functions are chosen homogeneous of degree 0:
\begin{equation}
\label{c2}
\begin{split}
&f\left(\frac{|\overrightarrow{p}|}{E}\right)=\sum_{k=1}^{\infty}\alpha_{k}\left(\frac{|\overrightarrow{p}|}{E}\right)^k\\ &g\left(\frac{\overrightarrow{p}}{E}\right)=\sum_{k=1}^{\infty}\beta_{k}\left(\frac{\overrightarrow{p}}{E}\right)^k
\end{split}
\end{equation}
The Modified Dispersion Relations result defined via a Finsler pseudo-norm $F(p)$, in fact the perturbation function homogeneity hypotheses permits to write the MDRs (\ref{c1}) as:
\begin{equation}
\label{c3}
\begin{split}
&MDR(p)=F^2(p)=m^2\\
&F(p)=\sqrt{E^2-\left(1-f\left(\frac{|\overrightarrow{p}|}{E}\right)-g\left(\frac{\overrightarrow{p}}{E}\right)\right)|\overrightarrow{p}|^2}
\end{split}
\end{equation}
Imposing the 0-degree homogeneity to the perturbation $f$ and $g$, one obtains that the function $F$, defined in (\ref{c3}), results homogeneous of degree 1, condition to be a candidate Finsler pseudo-norm. Important to underline the difference between a Finsler structure and a pseudo-Finsler one. The first geometric structure is constructed using a positively defined metric to pose the norm, instead the second one resorts to a not positive one. Here the pseudo-Finsler geometry is used, because one is dealing with the space-time structure, where the underlying global metric is the Minkowski one ($\eta_{\mu\nu}$), with signature $\{+,\,-,\,-,\,-\}$.\\
From here on, the perturbation function $g$ is posed equal to zero $(g=0)$ in order to try to construct an isotropic LIV theory. Hence only MDR, preserving rotational symmetry, will be considered:
\begin{equation}
\label{c4}
\begin{split}
&MDR(p)=F^2(p)=m^2\\
&F(p)=\sqrt{E^2-\left(1-f\left(\frac{|\overrightarrow{p}|}{E}\right)\right)|\overrightarrow{p}|^2}
\end{split}
\end{equation}
It is important to underline that in literature the form of the MDRs is usually:
\begin{equation}
\label{c5}
MDR(p)=E^2-\left(1-h(p)\right)|\overrightarrow{p}|^2=m^2
\end{equation}
with the perturbation $h(p)=\sum_{k=1}^{\infty}a_{k}\left(\frac{p}{M_{Pl}}\right)^{k}$, where $M_{Pl}$ represents the Planck mass, that is a suppression scale for the series expansion. The perturbation form introduced in this work can be justified because it is, for example, a subcase obtained in \cite{Koste4}, so it constitutes an interesting physical eventuality per se. Moreover, it is possible to demonstrate that every Modified Dispersion Relation of the form (\ref{c5}) can be approximated with a MDR of type (\ref{c4}), as in \cite{Torri,Antonelli}. In fact, perturbation function $f$, that preserves rotational symmetry can be rewritten in the form:
\begin{equation}
\label{c6}
f\left(\frac{|\overrightarrow{p}|}{E}\right)=\sum_{k=1}^{\infty}\alpha_{k}\left(\frac{|\overrightarrow{p}|}{E}\right)^k\Rightarrow f(\xi)=\sum_{k=1}^{\infty}\alpha_{k}\xi^k
\end{equation}
where $\xi=\frac{|\overrightarrow{p}|}{E}$.\\
To obtain the same physical effects description in both cases of MDR choice, it results necessary to impose the equality of (\ref{c4}) and (\ref{c5}), hence the following equation:
\begin{equation}
\label{c7}
f(\xi)=\sum_{k=1}^{\infty}\alpha_{k}\xi^k=h(p(\xi))
\end{equation}
From this equation it is possible to obtain a series expansion for $p$ as a function of $\xi$, that permits to fix the expansion coefficients $\alpha_{k}$ in order to satisfy the equation itself. In literature, for most physical cases, the $h(p)$ series terminates at first or second order. In these eventualities, it is always possible to find an approximation series for $p(\xi)$. This procedure is analogous to map a function of the variable $p=|\overrightarrow{p}|$ on a function of the variable $\xi$, noting that $\xi$ and $p$ have a biunivocal correspondence.\\
In order to verify that a dispersion relation defined with the introduction of the perturbation $f$ is well posed, it is necessary to verify that the energy solution of the equation, obtained using (\ref{c2}) and (\ref{c1}):
\begin{equation}
\label{c8}
E^2=p^2\left(1-\sum_{k=1}^{n}\alpha_{k}\left(\frac{p}{E}\right)^{k}\right)+m^2
\end{equation}
is positive for every $n$ value. Dividing the previous equation by $E^{2}$, it becomes:
\begin{equation}
\label{c9}
1=\left(\frac{p}{E}\right)^{2}\left(1-\sum_{k=1}^{n}\alpha_{k}\left(\frac{p}{E}\right)^{k}\right)+\frac{m^2}{E^2}
\end{equation}
in the very high energy scenario, that is taking the limit for $E\rightarrow\infty$, one obtains $\frac{m^2}{E^2}\rightarrow0$ and finally the equation (\ref{c9}) takes the form:
\begin{equation}
\label{c10}
1=\left(\frac{p}{E}\right)^{2}\left(1-\sum_{k=1}^{n}\alpha_{k}\left(\frac{p}{E}\right)^{k}\right)
\end{equation}
resorting again to the variable $\xi$ the previous relation becomes:
\begin{equation}
\label{c11}
\xi^{2}\left(1-P_{n}(\xi)\right)=1
\end{equation}
where $P_{n}(\xi)$ is a $n$ degree polynomial. If the magnitude of this polynomial remains limited, that is this function represents a tiny perturbation, compared to the magnitude of $p$, the solution of the equation is $\xi\simeq1$. So, posing the correct constrain on the coefficients of the series (\ref{c2}), one can obtain a real value energy $E$ from equation (\ref{c9}) and:
\begin{equation}
\label{c12}
\lim_{p\rightarrow\infty}\frac{p}{E}=1+\delta
\end{equation}
for a tiny positive constant $\delta$. Therefore it is possible to adopt homogeneous perturbation functions, under appropriate general assumptions.\\
Now, fixed the MDR general form it is possible to investigate the space-time induced modified geometrical structure.

\section{The Finsler geometric structure of space-time}
From the momentum space Finsler pseudo-norm, it is possible to determine the metric of the same space. Resorting to the hamiltonian formalism one can construct the space-time structure, starting from the modified momentum space geometry. This approach is compatible with \cite{Barcaroli}, where starting from the modified momentum geometry, the Hamiltonian is built. The explicit metric form, defined in momentum space, is obtained using the relation:
\begin{equation}
\label{c13}
\widetilde{g}(p)^{\mu\nu}=\frac{1}{2}\frac{\partial}{\partial p_{\mu}}\frac{\partial}{\partial p_{\nu}}F^2(E,\,\overrightarrow{p})
\end{equation}
It remains a non-diagonal part, which does not give any contribution in computing the dispersion relations. Therefore it can be eliminated by an opportune "gauge" choice \cite{Torri}. The final form of the metric becomes therefore:
\begin{equation}
\label{c14}
\widetilde{g}^{\mu\nu}(p)=\left(
                             \begin{array}{cc}
                                1 & 0 \\
                                0 & -(1-f(|\overrightarrow{p}|/E))\mathbb{I}_{3\times3} \\
                             \end{array}
                          \right)
\end{equation}
Consistently with standard relativity, the free massive particle \emph{Hamiltonian} is defined using the modified metric (\ref{c14}), as:
\begin{equation}
\label{c15}
\mathcal{H}=\sqrt{\widetilde{g}(p)^{\mu\nu}p_{\mu}p_{\nu}}=F(p)=MDR(p)
\end{equation}
written using the MDR, that is a pseudo-Finsler norm. Starting from this function, it is possible to compute the velocity, correlated to the momentum, employing the \emph{Legendre} transformation as:
\begin{equation}
\label{c16}
\dot{x}^{\mu}=\frac{\partial}{\partial p_{\mu}}F(p)\simeq\frac{\widetilde{g}(p)^{\mu\nu}\,p_{\nu}}{\sqrt{\widetilde{g}(p)^{\alpha\beta}\,p_{\alpha}\,p_{\beta}}}
\end{equation}
The homogeneity of the metric is fundamental to obtain this last equation, in fact it can be used to neglect the metric derivative by the momentum, which is proportional to terms like:
\begin{equation}
\label{c17}
\frac{\partial}{\partial p_{\mu}}f\left(\frac{|\overrightarrow{p}|}{E}\right)
\end{equation}
These terms can be neglected, because of the form of the perturbation function (\ref{c2}), that is, at high energies:
\begin{equation}
\label{c18}
\begin{split}
&\partial_{p^{j}}f(p)=\partial_{p^{j}}\sum_{k}\alpha_{k}\frac{|\overrightarrow{p}|^{k}}{E^{k}}\simeq\partial_{p^{j}}\sum_{k}\alpha_{k}\frac{|\overrightarrow{p}|^{k}}{(\sqrt{|\overrightarrow{p}|^2+m^2})^k}=\\
=&\sum_{k}\left(\alpha_{k}k\frac{|\overrightarrow{p}|^{k-2}p_{j}}{(\sqrt{|\overrightarrow{p}|^2+m^2})^k}-\alpha_{k}k\frac{|\overrightarrow{p}|^{k}p_{j}}{(\sqrt{|\overrightarrow{p}|^2+m^2})^{k+2}}\right)\rightarrow0
\end{split}
\end{equation}
where it has been used the equivalence $E\simeq\sqrt{|\overrightarrow{p}|^2+m^2}$.\\
Now it is possible to express the pseudo-Finsler norm as a function of coordinates:
\begin{equation}
\label{c19}
G(\dot{x}(p))=F(p)
\end{equation}
and the associated metric is given by the relation:
\begin{equation}
\label{c20}
g(x,\,\dot{x}(p))_{\mu\nu}=\frac{1}{2}\left(\frac{\partial^2G^2}{\partial\dot{x}^{\mu}\,\partial\dot{x}^{\nu}}\right)
\end{equation}
where $g_{\mu\nu}$ is the inverse of the previous (\ref{c15}) metric:
\begin{equation}
\label{c21}
g(x,\,\dot{x}(p))_{\mu\alpha}\;\widetilde{g}(x,\,p)^{\alpha\nu}=\delta_{\mu}^{\;\nu}
\end{equation}
and can be written as:
\begin{equation}
\label{c22}
g(x,\,\dot{x}(p))_{\mu\nu}=\left(
                     \begin{array}{cc}
                         1 & 0 \\
                         0 & -\frac{\mathbb{I}_{3\times3}}{(1-f(|\overrightarrow{p}|/E))} \\
                     \end{array}
                  \right)\\=
\left(
    \begin{array}{cc}
       1 & 0 \\
       0 & -(1+f(|\overrightarrow{p}|/E))\mathbb{I}_{3\times3} \\
    \end{array}
\right)
\end{equation}
Starting from the Hamiltonian (\ref{c15}), it is possible to compute the explicit form of the \emph{Lagrangian}, that is:
\begin{equation}
\label{c23}
\begin{split}
\mathcal{L}=&\overrightarrow{p}\,\overrightarrow{\dot{x}}-\mathcal{H}=-\dot{x}^{\mu}\,p_{\mu}=\left(\frac{\partial}{\partial \dot{x}^{\mu}}\mathcal{L}\right)\,\dot{x}^{\mu}=-m\sqrt{\dot{x}^{\mu}\,g_{\mu\nu}(p)\,\dot{x}^{\nu}}
\end{split}
\end{equation}
The geometric structure of the obtained space-time permits to preserve the Hamilton-Jacobi equations structure. In fact the momentum takes the explicit form:
\begin{equation}
\label{c24}
p_{\mu}=\left(-\frac{\partial}{\partial \dot{x}^{\mu}}\mathcal{L}\right)=\,\frac{m\,g_{\mu\nu}\,\dot{x}^{\nu}}{\sqrt{\dot{x}^{\tau}\,g_{\tau\sigma}(p)\,\dot{x}^{\sigma}}}
\end{equation}
where again the homogeneity of the metric $g_{\mu\nu}$ (\ref{c22}) has been used to justify the neglecting of the metric derivative. The momentum satisfies the mass-shell condition:
\begin{equation}
\label{c25}
\begin{split}
\widetilde{g}^{\mu\nu}(p)\,p_{\mu}\,p_{\nu}\simeq&\,\widetilde{g}^{\mu\nu}\,\frac{m\,g_{\mu\alpha}(p)\,\dot{x}^{\alpha}}{\sqrt{\dot{x}^{\tau}\,g_{\tau\sigma}(p)\,\dot{x}^{\sigma}}}\,\frac{m\,g_{\nu\beta}(p)\,\dot{x}^{\beta}}{\sqrt{\dot{x}^{\tau}\,g_{\tau\sigma}(p)\,\dot{x}^{\sigma}}}=\\
=&\,g_{\mu\nu}(p)\dot{x}^{\mu}\dot{x}^{\nu}=m^2
\end{split}
\end{equation}
that is the MDR relation (\ref{c4}).\\
Now it is useful to deal with the obtained pseudo-Finsler metric structure of the space-time, introducing the Cartan formalism, that is resorting to the \emph{vierbein} or \emph{tetrad}. In this work the vierbein acquires an explicit dependence on energy-momentum, because one is dealing with an energy depending Finsler pseudonorm defined space-time.\\
Remembering that two \emph{vierbein} are equivalent if they originate the same metric:\footnote{A simple example is $e'^{a}_{\;\mu}(x)=-e^{a}_{\;\mu}(x)$.}:
\begin{equation}
\label{c26}
e^{a}_{\;\mu}(x)\,\eta_{ab}\,e^{b}_{\;\nu}(x)=g_{\mu\nu}(x)=e'^{a}_{\;\mu}(x)\,\eta_{ab}\,e'^{b}_{\;\nu}(x)
\end{equation}
from now on \emph{thetrad} elements equivalence classes will be considered, identifying every class with one representative.
To originate the (\ref{c14}) metric, the \emph{vierbein} must have the following expression:
\begin{equation}
\label{c27}
\begin{split}
& e^{\mu}_{\,a}(p)=        \left(
                             \begin{array}{cc}
                               1 & \overrightarrow{0} \\
                               \overrightarrow{0}^{t} & \sqrt{1-f(p)}\,\mathbb{I}_{3\times3} \\
                             \end{array}
                           \right)\\ \\
& e_{\mu}^{\,a}(p)=        \left(
                             \begin{array}{cc}
                               1 & \overrightarrow{0} \\
                               \overrightarrow{0}^{t} & \frac{\mathbb{I}_{3\times3}}{\sqrt{1-f(p)}} \\
                             \end{array}
                           \right)\simeq
                           \left(
                             \begin{array}{cc}
                               1 & \overrightarrow{0} \\
                               \overrightarrow{0}^{t} & \sqrt{1+f(p)}\,\mathbb{I}_{3\times3} \\
                             \end{array}
                           \right)
\end{split}
\end{equation}
where the explicit dependence on the momentum magnitude has been intoduced.
To analyze the geometric structure it is now necessary to introduce the \emph{affine} connection:
\begin{equation}
\label{c28}
\Gamma_{\mu\nu}^{\,\alpha}=\frac{1}{2}g^{\alpha\beta}\left(\partial_{\mu}g_{\beta\nu}+\partial_{\nu}g_{\mu\beta}-\partial_{\beta}g_{\mu\nu}\right)
\end{equation}
Following \cite{Torri} and \cite{Antonelli}, it is simple to evaluate the explicit forms of the affine connection components, starting from the metric tensor. The result is identically equal to zero for the following Christoffel symbols:
\begin{equation}
\label{c29}
\Gamma_{\mu0}^{\,0}=\Gamma_{00}^{\,i}=\Gamma_{\mu\nu}^{\,i}=0\qquad\forall \mu\neq\nu
\end{equation}
Even the not null components can be approximated by zero:
\begin{equation}
\label{c29a}
\begin{split}
&\Gamma_{ii}^{\,0}=-\frac{1}{2}\partial_{0}f(p)\simeq0\\
&\Gamma_{0i}^{\,0}=\Gamma_{i0}^{\,0}=\frac{1}{2(1+f(p))}\partial_{0}f(p)\simeq0\\
&\Gamma_{ii}^{\,i}=\frac{1}{2(1+f(p))}\partial_{i}f(p)\simeq0\\
&\Gamma_{jj}^{\,i}=-\frac{1}{2(1+f(p))}\partial_{i}f(p)\simeq0\qquad\forall i\neq j\\
&\Gamma_{ij}^{\,i}=\Gamma_{ji}^{\,i}=\frac{1}{2(1+f(p))}\partial_{i}f(p)\simeq0\qquad\forall i\neq j
\end{split}
\end{equation}
since the derivative $|\partial_{p}f(p)|$ can be neglected. This results possible under the assumption of tiny interaction with the space-time background structure and thanks to the homogeneity of the perturbation functions (\ref{c2}) and to equations (\ref{c18}). In previous equations, latin indices indicate spatial tensor components (they belong to the values set $\{1,\,2,\,3\}$), greek ones instead indicate all the four space-time components (they variate inside the set $\{0,\,1,\,2,\,3\}$), usual convention of General Relativity.\\
Introducing the local covariant derivative as:
\begin{equation}
\label{c30}
\nabla_{\mu}v^{\nu}=\partial_{\mu}v^{\nu}+\Gamma_{\mu\alpha}^{\,\nu}v^{\alpha}\simeq\partial_{\mu}v^{\nu}
\end{equation}
it is immediate to compute the \emph{Cartan} or \emph{spinorial} connection, as:
\begin{equation}
\label{c31}
\omega_{\mu ab}=e^{\;\nu}_{a}\nabla_{\mu}e_{b\nu}\simeq e^{\;\nu}_{a}\partial_{\mu}e_{b\nu}
\end{equation}
Applying the first Cartan structural equation in differential form:
\begin{equation}
\label{c32}
de=e\wedge\omega
\end{equation}
to the external forms
\begin{equation}
\label{c33}
e_{0}^{\,\mu}=dx^{\mu}\qquad e_{i}^{\,\mu}=\sqrt{1-f(p)}dx^{\mu}
\end{equation}
it follows that even for the spinorial connection, the not null elements are negligible:
\begin{equation}
\label{c34}
\frac{1}{2}\epsilon_{ijk}\omega^{ij}=\frac{1}{2}\frac{1}{1-f}\epsilon_{ijk}(\partial^{i}fdx^{j}-\partial^{j}fdx^{i})\simeq0
\end{equation}
because, as in the previous case, they are proportional to perturbation functions derivatives.
So, even for the Cartan connection, the not null coefficients are proportional to terms like (\ref{c18}), hence, the connection is asymptotically zero $(\omega_{\mu ab}\simeq0)$. The tensor total covariant derivative results therefore:
\begin{equation}
\label{c36}
D_{\mu}v^{\;\nu}_{a}=\partial_{\mu}v^{\;\nu}_{a}+\Gamma_{\mu\alpha}^{\,\nu}v^{\;\alpha}_{a}-\omega_{\mu\nu}^{\,a}v^{\;\nu}_{b}\simeq\partial_{\mu}v^{\;\nu}_{b}
\end{equation}
At the end it is possible to conclude that the supposed massive particle tiny interaction with the ``quantized" space-time background, determines an asymptotically flat Finslerian structure.

\section{Modified Lorentz Transformations}
Using the \emph{tetrad}, it is possible to construct the explicit form of the modified Lorentz group. The obtained representation preserves the form of the MDR and the homogeneity of degree $0$ of the perturbation functions.\\
In literature \cite{Carroll} it is possible to find the general form of the Lorentz transformations for General Relativity, defined as:
\begin{equation}
\label{c37}
\Lambda_{\mu}^{\;\nu}(x)e_{\nu}(x)=e_{\mu}(\Lambda x)
\end{equation}
Resorting to the \emph{vierbein} it is possible to define projection from a tangent (local) space, parameterized by the metric $g_{\mu\nu}(x,\,v)$ to another local space, identified by a different metric tensor $\overline{g}(x',\,v')_{\mu\nu}$ as summarized in the following graph:
\[\begin{tikzcd}
         (TM,\,\eta_{ab},\,v) \arrow{d}{e(x)} \arrow{rr}[swap]{\Lambda} && (TM,\,\eta_{ab},\,v')\arrow{d}[swap]{\overline{e}(x')} \\
        (T_{x}M,\,g_{\mu\nu}(x),v) \arrow{rr}[swap]{\overline{e}\circ\Lambda\circ e^{-1}} && (T_{x}M,\,\overline{g}_{\mu\nu}(x'),v')
\end{tikzcd}\]
From now on, the dependence of thetrad and the metric tensor will be generalized from the space-time coordinates $(x)$ to the coordinates of the phase space $(x,\,p)$. In this way a dependence on the momentum is included, like in Finsler geometry \cite{Chern1}. The dependence on the position is supposed trivial \cite{Torri,Antonelli} and therefore will be neglected to preserve the space homogeneity. Only the dependence on momentum (velocity) is maintained and all the physical quantities are generalized, acquiring an explicit dependence on it. The graph of the transition from one tangent (local) space to the other becomes:
\[\begin{tikzcd}
         (TM,\,\eta_{ab},\,p) \arrow{d}{e(p)} \arrow{rr}[swap]{\Lambda} && (TM,\,\eta_{ab},\,p')\arrow{d}[swap]{\overline{e}(p')} \\
        (T_{x}M,\,g_{\mu\nu}(p)) \arrow{rr}[swap]{\overline{e}\circ\Lambda\circ e^{-1}} && (T_{x}M,\,\overline{g}_{\mu\nu}(p'))
\end{tikzcd}\]
where is indicated the explicit dependence of the metric from momentum.\\
Using the vierbein to transform a latin (global index) in a greek one (local index), it is possible to write:
\begin{equation}
\label{c38}
g_{\mu\nu}(p)=e^{a}_{\;\mu}(p)\,\eta_{ab}\,e^{b}_{\nu}(p)=e^{a}_{\;\mu}(p)\,\Lambda_{a}^{\;c}\,\eta_{cd}\,\Lambda_{b}^{\;d}\,e^{b}_{\nu}(p)
\end{equation}
From the previous equation it follows:
\begin{equation}
\label{c39}
\Lambda_{a}^{\;c}e^{a}_{\;\mu}(p)=e^{c}_{\;\mu}(p)
\end{equation}
This result permits to correlate the global Lorentz transformations with the vierbein elements, indicating how a tetrad element transform under such transformations.\\
Now, using again the vierbein to transform local to global indices, it is possible to define the general modified Lorentz transformation as:
\begin{equation}
\label{c40}
\Lambda_{\mu}^{\;\nu}(p,\,\Lambda p)=e_{\;\mu}^{a}\,(\Lambda p)\Lambda_{a}^{\;b}\,e^{\;\nu}_{b}(p)
\end{equation}
Using the equation (\ref{c37}), with the substitution of coordinate with momentum, and equation (\ref{c40}), it is possible to write:
\begin{equation}
\label{c41}
\Lambda_{\mu}^{\;\nu}(p,\,\Lambda p)e^{a}_{\;\nu}(p)=\underbrace{e_{\mu}^{\;b}(\Lambda p)\Lambda_{b}^{\;c}e_{c}^{\;\nu}(p)}_{\Lambda_{\mu}^{\;\nu}(p,\,\Lambda p)}e^{a}_{\;\nu}(p)=e_{\mu}^{\;b}(\Lambda p)\delta_{b}^{\;a}=e_{\mu}^{\;a}(\Lambda p)
\end{equation}
where in the last equality relation (\ref{c39}) has been used.\\
Considering the MDR (\ref{c4}) and remembering the momentum space metric is given by (\ref{c14}), it is now possible to verify that the Modified Lorentz Transformations (MLT) are isometries for the Modified Dispersion Relation, that is $MDR(\Lambda p)=MDR(p)$. In fact:
\begin{equation}
\begin{split}
\label{c42}
&MDR(\Lambda p)=\Lambda_{\mu}^{\;\alpha}(p,\,\Lambda p)p_{\alpha}g^{\mu\nu}(\Lambda p)\Lambda_{\nu}^{\;\beta}(p,\,\Lambda p)p_{\beta}=\underbrace{e_{\;\mu}^{a}(\Lambda p)\Lambda_{a}^{\;b}e_{b}^{\;\alpha}(p)}_{\Lambda_{\mu}^{\;\alpha}(p,\,\Lambda p)}\,p_{\alpha}\\& g^{\mu\nu}(\Lambda p)\underbrace{e_{\;\nu}^{c}(\Lambda p)\Lambda_{c}^{\;d}e_{d}^{\;\beta}(p)}_{\Lambda_{\nu}^{\;\beta}(p,\,\Lambda p)}\,p_{\beta}=e_{\;\mu}^{a}(\Lambda p)\Lambda_{a}^{\;b}\,p_{b}\,g^{\mu\nu}(\Lambda p)\,e_{\;\nu}^{c}(\Lambda p)\Lambda_{c}^{\;d}p_{d}=\\
=&\;\Lambda_{a}^{\;b}p_{b}\,\eta^{ac}\,\Lambda_{c}^{\;d}\,p_{d}=p_{a}\eta^{ab}\,p_{b}=e^{a}_{\;\mu}(p)\,p_{a}\,g^{\mu\nu}(p)e^{b}_{\;\nu}(p)\,p_{b}=MDR(p)
\end{split}
\end{equation}
where the equalities (\ref{c40}) and (\ref{c41}), obtained before, have been used.\\
The Modified Lorentz Transformations, introduced in (\ref{c40}), are therefore the isometries of the Modified Dispersion Relations.\\
Moreover the amended Lorentz group transformations (\ref{c40}) acting on the 4-vector $p^{\mu}=(E,\,\overrightarrow{p})$, give, for the modification function $f$ in the MDR:
\begin{equation}
\label{c43}
f\left(\frac{|\overrightarrow{p}|}{E}\right)\rightarrow f\left(\frac{|\Lambda^{i}_{\;\mu}(p)p^{\mu}|}{\Lambda^{0}_{\;\mu}(p)p^{\mu}}\right)
\end{equation}
It is simple to verify that this kind of transformations preserve the homogeneity of degree 0, because of the ratio present in the definition of the modification function $f$. Therefore the action of the modified Lorentz group preserves the homogeneity of the perturbation function $f$, preserving the MDR form (\ref{c4}).

\section{Very Special Relativity (VSR) correspondence}
MDRs can be generalized in a form, which includes energy dependent corrections, as for example in \cite{Smolin1,Smolin2}:
\begin{equation}
\label{c106}
f_{1}^{2}E^2-f_{2}^{2} |\overrightarrow{p}|^2=m^2\, ,
\end{equation}
where $f_{i}$ are four-momentum $p$ functions. These functions can be written in a perturbative fashion as $f_{i}=1-h_{i}$, where $h_{i}\ll1$ are the velocities modification parameters. From this relation, it is possible to derive an explicit equality for the energy:
\begin{equation}
\label{c107}
E=\sqrt{\frac{m^2}{f_{1}^{2}}+\frac{f_{2}^{2}}{f_{1}^{2}}\,|\overrightarrow{p}|^2}\simeq p f_{3}\,\,,with\,f_{3}=\frac{f_{2}}{f_{1}}
\end{equation}
The velocity of the particle can be obtained using Hamilton-Jacobi equation:
\begin{equation}
\label{c108}
c'(E)=\frac{\partial}{\partial p}E\bigg\vert_{max}=(f_{(3)}+p\,f_{(3)}')=\left(f_{(3)}+f_{(3)}'\,p\,\left(\frac{1}{E}-\frac{p}{E^2}\right)\right)
\end{equation}
Therefore every massive particle feels a local space-time foliation, depending on its momentum. From this the necessity follows to resort to Finsler geometry, that can deal with this local space-time momentum depending parametrization.\\
Returning now to HMSR model, where $f_{1}^{\;2}=1$, $f_{2}^{\;2}=(1-f)$ and $f$ is the homogeneous perturbation, introduced in this work (\ref{c2}), if its  magnitude remains negligible, compared to the momentum, the ratio $\frac{|\overrightarrow{p}|}{E}$ have a finite limit for $p\longrightarrow\infty$ ($\frac{|\overrightarrow{p}|}{E}\longrightarrow1+\delta$). Consequently, even the $f$ function admits finite limit, $f(1+\delta)\longrightarrow\epsilon$. In this way the perturbation $f_3$, for $p\longrightarrow\infty$, tends to
$ \lim_{p\longrightarrow\infty} \, f_3^{\;2} = 1- f(1+\delta) = 1-\epsilon\, $.  Therefore it is possible to obtain the Coleman and Glashow's ``Very Special Relativity'' (VRS) scenario as a high energy (high momentum) limit \cite{Glash1,Glash2}. In this case it is possible to recover from equation (\ref{c108}) a massive particle ``personal'' \emph{maximum attainable velocity} $c'$:
\begin{equation}
\label{c109}
c'(E)=f_{3}=\sqrt{1-\epsilon}
\end{equation}
because $f'_{(3)}\rightarrow0$ for $p\longrightarrow \infty$, reobtaining a result provided in \cite{Glash1}.\\
It is also possible to show that the modified Lorentz group, introduced in HMSR, is compatible with the special relativity transformations computed introducing a personal maximum attainable velocity, different for every particle species. It is well known that this corresponds to ignore the speed of light universality postulate, in computing the Lorentz transformations \cite{Ignat,Ignat1,Ignat2,Ignat3,Jin}. In fact, supposing the case of Lorentz boost along the "x" direction, the explicit form of the Modified Lorentz Transformations (MLT) is given, using equation (\ref{c40}), by:
\begin{equation}
\label{c110}
\begin{split}
&\Lambda^{\mu}_{\,\nu}(p',\,p)=e_{a}^{\,\mu}(p')\Lambda^{a}_{\,b}e^{b}_{\,\nu}(p)=e_{a}^{\,\mu}(\Lambda p)\Lambda^{a}_{\,b}e^{b}_{\,\nu}(p)=\\
=&\left(
    \begin{array}{cccc}
      1 & 0 & 0 & 0 \\
      0 & \chi & 0 & 0 \\
      0 & 0 & \chi & 0 \\
      0 & 0 & 0 & \chi \\
    \end{array}
  \right)
  \left(
    \begin{array}{cccc}
      \gamma & -\beta\,\gamma & 0 & 0 \\
      -\beta\,\gamma & \gamma & 0 & 0 \\
      0 & 0 & 1 & 0 \\
      0 & 0 & 0 & 1 \\
    \end{array}
  \right)
  \left(
    \begin{array}{cccc}
      1 & 0 & 0 & 0 \\
      0 & 1/\xi & 0 & 0 \\
      0 & 0 & 1/\xi & 0 \\
      0 & 0 & 0 & 1/\xi \\
    \end{array}
  \right)=\\
 =&\left(
    \begin{array}{cccc}
      \gamma & -\beta/\xi\,\gamma & 0 & 0 \\
      -\beta\,\chi\gamma & \chi/\xi\,\gamma & 0 & 0 \\
      0 & 0 & \chi/\xi & 0 \\
      0 & 0 & 0 & \chi/\xi \\
    \end{array}
  \right)\simeq
  \left(
    \begin{array}{cccc}
      \gamma & -\beta/\xi\,\gamma & 0 & 0 \\
      -\beta\,\chi\gamma & \chi/\xi\,\gamma & 0 & 0 \\
      0 & 0 & 1 & 0 \\
      0 & 0 & 0 & 1 \\
    \end{array}
  \right)
\end{split}
\end{equation}
where $\beta=v/c$, $\gamma=1/\sqrt{1-\beta^{2}}$ and $\chi=\sqrt{1-f(p')}$ and $\xi=\sqrt{1-f(p)}$. The ratio $\chi/\xi$ in the last two diagonal terms can be approximated with $1$, because this term corrections are negligible, compared with the other matrix coefficients.\\
The transformations obtained are correlated with the natural coordinate units of measure and these MLT are valid for coordinates $(ct,\,\overrightarrow{x})$. The maximum attainable velocity in the two reference frames, denoted by the momenta $p$ and $p'$ are:
\begin{equation}
\label{c111}
\begin{cases}
\begin{split}
&c(p')=\chi(p')\,c_{0}\\
&c(p)=\xi(p)\,c_{0}
\end{split}
\end{cases}
\end{equation}
where $c_{0}$ is a fixed constant value and $c_{0}=1$ in natural measure units.\\
To convert these MLT to the usual coordinates $\{t,\,x,\,y,\,z\}$, it is necessary to determine the value of $x'$ in the transformed reference frame, noting that:
\begin{equation}
\label{c112}
\gamma'\simeq\frac{\chi}{\xi}\,\gamma
\end{equation}
This is compatible with the form of the coefficients:
\begin{equation}
\label{c113}
\gamma=\frac{c(p)}{\sqrt{c(p)^2-v^2}}\qquad \gamma'=\frac{c(p')}{\sqrt{c(p')^2-v'^2}}
\end{equation}
from which it follows that equation (\ref{c112}) is correct. The final MLT form for the usual standard coordinates $\{t,\,x,\,y,\,z\}$ results therefore:
\begin{equation}
\label{c114}
\Lambda^{\mu}_{\;\nu}(p,\,\Lambda p)=
\left(
    \begin{array}{cccc}
      (\chi/\xi)\,\gamma & -v\,(\xi/\chi)\,\gamma & 0 & 0 \\
      -v\,(\chi/\xi)\,\gamma & (\chi/\xi)\,\gamma & 0 & 0 \\
      0 & 0 & 1 & 0 \\
      0 & 0 & 0 & 1 \\
    \end{array}
  \right)
\end{equation}
where the first term of the first row has been multiplied by the ratio $c(p')/c(p)$ and the second term of the first row has been divided by $c^2(p)$, to convert the measure units. These results are compatible with special relativity constructed without the light speed postulate, that is with personal maximum attainable velocities \cite{Jin}, therefore the construction here introduced is coherent.

\section{Relativistic Invariant Energy (Mandelstam variables)}
In HMSR every particle species has its own metric, with a personal maximum attainable velocity. Moreover every species presents its personal Modified Lorentz Transformations (MLT), which are the isometries for the Modified Dispersion Relation (MDR) of the particle. The new physics, caused by LIV, emerges only in the interaction of two different species. That is every particle type physics is modified in a different way by the Lorentz symmetry violation. Therefore, to analyze the interaction of two particles, it is necessary to determine how the reaction invariants - that is the Mandelstam relativistic invariants - are modified.\\
Starting from the hypothesis of MDR, generated by a metric in the momentum space, it is necessary to resort to the \emph{vierbein} to project the particles momenta on the Minkowski tangent space:
\begin{equation}
\label{c44}
MDR(p)=p_{\mu}\,g^{\mu\nu}(p)\,p_{\nu}=p_{\mu}\,e_{a}^{\;\mu}(p)\,\eta^{ab}\,e_{b}^{\;\nu}(p)\,p_{\nu}=p_{a}\,\eta^{ab}\,p_{b}
\end{equation}
For this reason it seems natural to generalize the definition of internal product of the sum of two different particle species momenta as:
\begin{equation}
\label{c45}
\langle p+q|p+q \rangle=(p_{\mu}\,e_{a}^{\,\mu}(p)+q_{\mu}\,\tilde{e}_{a}^{\,\mu}(q))\,\eta^{ab}\,(p_{\nu}\,e_{b}^{\,\nu}(p)+q_{\nu}\,\tilde{e}_{b}^{\,\nu}(q))
\end{equation}
where $e$ indicates the tetrad related to the first particle and with $\tilde{e}$ the vierbein related to the second one. With this internal product it is now possible to define the Mandelstam variables $s$, $t$ and $u$, remembering that:
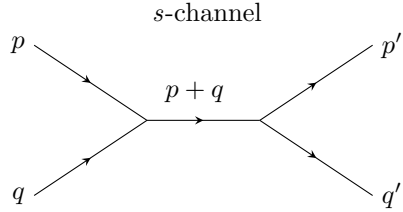
\tikzset{
particle/.style={thick,draw=black, postaction={decorate},
    decoration={markings,mark=at position .5 with {\arrow[black]{triangle 45}}}},
gluon/.style={decorate, draw=black,
    decoration={coil,aspect=0}}
 }
\tikzset{
particle/.style={thin,draw=black, postaction={decorate},
decoration={markings,mark=at position .5 with {\arrow[black]{stealth}}}},
gluon/.style={decorate, draw=black, decoration={snake=coil}}
}
\begin{center}
\emph{s}-channel\\
\begin{tikzpicture}[node distance=1cm and 1.5cm]
\coordinate[label=left:$p$] (e1);
\coordinate[below right=of e1] (aux1);
\coordinate[right=of aux1] (aux2);
\coordinate[above right=of aux2,label=right:$p'$] (e2);
\coordinate[below left=of aux1,label=left:$q$] (e3);
\coordinate[below right=of aux2,label=right:$q'$] (e4);

\draw[particle] (e1) -- (aux1);
\draw[particle] (aux2) -- (e2);
\draw[particle] (e3) -- (aux1);
\draw[particle] (aux2) -- (e4);
\draw[particle] (aux1)  node[label=above right:$p+q$] {} -- (aux2);
\end{tikzpicture}
\end{center}
\begin{center}
\emph{t}-channel\\

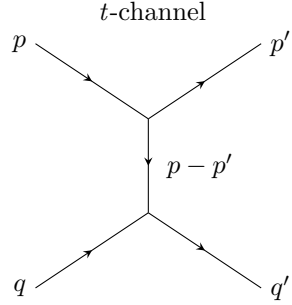
\begin{tikzpicture}[node distance=1cm and 1.5cm]
\coordinate[label=left:$p$] (e1);
\coordinate[below right=of e1] (aux1);
\coordinate[above right=of aux1,label=right:$p'$] (e2);
\coordinate[below=1.25cm of aux1] (aux2);
\coordinate[below left=of aux2,label=left:$q$] (e3);
\coordinate[below right=of aux2,label=right:$q'$] (e4);

\draw[particle] (e1) -- (aux1);
\draw[particle] (aux1) -- (e2);
\draw[particle] (e3) -- (aux2);
\draw[particle] (aux2) -- (e4);
\draw[particle] (aux1) -- node[label=right:$p-p'$] {}(aux2);
\end{tikzpicture}
\end{center}

\begin{center}
\emph{u}-channel\\
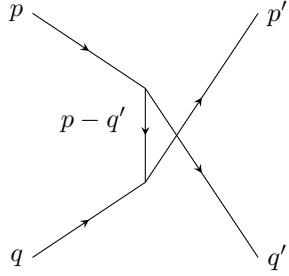
\begin{tikzpicture}[node distance=1cm and 1.5cm]
\coordinate[label=left:$p$] (e1);
\coordinate[below right=of e1] (aux1);
\coordinate[above right=of aux1,label=right:$p'$] (e2);
\coordinate[below=1.25cm of aux1] (aux2);
\coordinate[below left=of aux2,label=left:$q$] (e3);
\coordinate[below right=of aux2,label=right:$q'$] (e4);

\draw[particle] (e1) -- (aux1);
\draw[particle] (aux2) -- (e2);
\draw[particle] (e3) -- (aux2);
\draw[particle] (aux1) -- (e4);
\draw[particle] (aux1) node[label=below left:$p-q'$] {} -- (aux2);
\end{tikzpicture}
\end{center}
and considering the $p$ and $q$ momenta as belonging in general to different particle species.\\
If the two interacting particles belong to the same species, the internal product and therefore the Mandelstam variables present no differences from standard Physics. That is the momenta of particles of the same kind, live in the same tangent (local) space, constructed with the Finsler metric.  Instead, if the particles belong to different species, the definition of the internal product requires the necessity to correlate different local tangent spaces.\\
The new internal product can be generated introducing the concept of a generalized metric, written as:
\begin{equation}
\label{c46}
G=\left(
    \begin{array}{cc}
      g^{\mu\nu}(p) & e^{a\mu}(p)\tilde{e}_{a}^{\;\beta}(q) \\
      \tilde{e}^{a\alpha}(q)e_{a}^{\;\nu}(p) & \tilde{g}^{\alpha\beta}(q) \\
    \end{array}
  \right)
\end{equation}
The Modified Lorentz Transfomations for this metric assumes the explicit form:
\begin{equation}
\label{c47}
\Lambda=
\left(
  \begin{array}{cc}
    \Lambda_{\mu}^{\;\mu'} & 0 \\
    0 & \tilde{\Lambda}_{\alpha}^{\;\alpha'} \\
  \end{array}
\right)
\end{equation}
using the MLT of the two particle species.
The internal product (\ref{c45}) can be obtained as:
\begin{equation}
\begin{split}
\label{c48}
&\langle p+q|p+q \rangle=
\left(
  \begin{array}{cc}
    p & q \\
  \end{array}
\right)
\left(
    \begin{array}{cc}
      g^{\mu\nu}(p) & e^{a\mu}(p)\tilde{e}_{a}^{\;\beta}(q) \\
      \tilde{e}^{a\alpha}(q)e_{a}^{\;\nu}(p) & \tilde{g}^{\alpha\beta}(q) \\
    \end{array}
\right)
\left(
  \begin{array}{c}
    p \\
    q \\
  \end{array}
\right)=\\
=&p_{\mu}\,g^{\mu\nu}(p)\,p_{\nu}+p_{\mu}\, e^{a\mu}\tilde{e}_{a}^{\;\beta}(q)\,q_{\beta}+q_{\alpha}\,\tilde{e}^{a\alpha}(q)\,e_{a}^{\;\nu}(p)\,p_{\nu}+q_{\alpha}\tilde{g}^{\alpha\beta}(q)\,q_{\beta}
\end{split}
\end{equation}
The Mandelstam variables result covariant under the action of the new introduced MLT (\ref{c47}), in the same way as the MLT (\ref{c40}) are the isometries for the MDR for every particle species:
\begin{equation}
\begin{split}
\label{c49}
&\langle p+q|p+q \rangle=
\left(
  \begin{array}{cc}
    p & q \\
  \end{array}
\right)
\left(
    \begin{array}{cc}
      g^{\mu\nu}(p) & e_{a\mu}(p)\tilde{e}_{a}^{\;\beta}(q) \\
      \tilde{e}^{a\alpha}(q)e_{a}^{\;\nu}(p) & \tilde{g}^{\alpha\beta}(q) \\
    \end{array}
\right)
\left(
  \begin{array}{c}
    p \\
    q \\
  \end{array}
\right)=\\
=&\langle\Lambda(p+q)|\Lambda(p+q)\rangle=
\left[
\left(
  \begin{array}{cc}
    \Lambda_{\mu'}^{\;\mu} & 0 \\
    0 & \tilde{\Lambda}_{\alpha'}^{\;\alpha} \\
  \end{array}
\right)
\left(
\begin{array}{c}
  p \\
  q
\end{array}
\right)
\right]\cdot\\
&\left(
    \begin{array}{cc}
      g^{\mu'\nu'}(\Lambda p) & e^{a\mu'}(\Lambda p)\tilde{e}_{a}^{\;\beta'}(\tilde{\Lambda}q) \\
      \tilde{e}^{a\alpha}(\tilde{\Lambda}q)e_{a}^{\;\nu'}(\Lambda p) & \tilde{g}^{\alpha'\beta'}(\tilde{\Lambda}q) \\
    \end{array}
\right)\cdot
\left[\left(
  \begin{array}{cc}
    \Lambda_{\nu'}^{\;\nu} & 0 \\
    0 & \tilde{\Lambda}_{\beta'}^{\;\beta} \\
  \end{array}
\right)
\left(
\begin{array}{c}
  p \\
  q
\end{array}
\right)
\right]
\end{split}
\end{equation}
where equation (\ref{c40}) has been repeatedly used and the internal product is defined using the (\ref{c14}) momentum depending metric tensor.\\
The complete physical description of interactions can be made using the formalism of the $S$ matrix, which results to be an analytic function of the Madelstam variables. Since these quantities result covariant, respect to the amended Lorentz transformations (MLT), the concept of isotropy is restored. In this way the necessity of introducing a privileged class of inertial observers disappears.\\
In case of composition of three particles momenta, each of different type, the internal product (\ref{c45}) can be constructed by an analogous process. In fact it is possible to introduce a new metric, that integrates the different types of vierbeins and metrics relative to the three particles species:
\begin{equation}
\label{c50}
G=\left(
  \begin{array}{ccc}
    g^{\mu\nu}(p_{1}) & e^{a\mu}(p_{1})\widetilde{e}_{a}^{\;\beta}(p_{2}) & e^{a\mu}(p_{1})\overline{e}_{a}^{\;\rho}(p_{3}) \\
    \widetilde{e}^{a\alpha}(p_{2})e_{a}^{\;\nu}(p_{1}) & \widetilde{g}^{\alpha\beta}(p_{2}) & \widetilde{e}^{a\alpha}(p_{2})\overline{e}_{a}^{\;\rho}(p_{3}) \\
    \overline{e}^{a\theta}(p_{3})e_{a}^{\;\nu}(p_{1}) & \overline{g}^{a\theta}(p_{3})\widetilde{e}_{a}^{\;\beta}(p_{2}) & \overline{g}^{\theta\rho}(p_{3}) \\
  \end{array}
  \right)
\end{equation}
where $g$ and $e$ are related to the first particle species, $\widetilde{g}$ and $\widetilde{e}$ are related to the second one and $\overline{g}$ and $\overline{e}$ to the third one. All the processes, introduced for the two particles interaction, can be generalized in this way for generic n-particles (n-momenta) interactions. Finally the amended Lorentz group acquires the explicit form:
\begin{equation}
\label{c47a}
\Lambda_{\mu}^{\;\nu}(p,\,\Lambda p)=\otimes_{i}\Lambda^{(i)\;\nu}_{\;\;\;\mu}(p,\,\Lambda p)
\end{equation}
where the direct product is to be intended on the different particle species (i).

\section{Modified momenta composition rules and Double Special Relativity (DSR) correspondence}
In Double Special Relativity (or $\kappa$-deformed relativity) \cite{AmelinoCamelia} the starting point consists again in modifying the kinematics of the interaction processes, requiring the invariance of the formulation with respect to new introduced (modified) Lorentz transformations. To satisfy this requirement, in these theories the geometry of the momentum space is modified, introducing a \emph{modified composition rule} for the momenta:
\begin{equation}
\label{c115}
(p,\,q)\rightarrow(p\oplus q)=p+q+f(p,\,q)
\end{equation}
where $f(p,\,q)$ represents a perturbation of the usual momenta sum. At the same time the inverse operation is introduced, which allows to obtain incoming momenta from the outgoing ones: $(\ominus p) \oplus p=0$. These definitions correspond to the replacement of the momentum with a modified one, given by the relation:
\begin{equation}
\label{c116}
\pi_{\mu}=M_{\mu}^{\;\nu}(p)p_{\nu}
\end{equation}
with the transformations $M_{\mu}^{\;\nu}(p)$ determined by the geometric features of the momentum space \cite{Koste5}. The geometry of the momentum space can be determined from the algebraic properties generated by the modified composition rule \cite{AmelinoCamelia}, with the affine connection given by:
\begin{equation}
\label{c117}
\frac{\partial}{\partial p_{a}}\frac{\partial}{\partial q_{b}} (p\oplus q)|_{c}=\Gamma^{ab}_{c}
\end{equation}
This class of Relativity modification theories present the advantage of preserving covariance respect to the introduced Modified Lorentz Transformations. In HMSR an analogous idea is explored, modifying the momentum space geometry with the introduction of corrections of kinematical nature only in particle propagation, which can be compared to the modified composition rule of momenta. From the definition of the internal product (\ref{c45}) it is possible to recognize that the vierbein elements are used to project different momenta on a common support space. In fact the only space where two different species momenta can "live together" is the Minkowski one, that underlies all the personal spaces of every particle. Starting from this observation it is possible to obtain a modified composition rule for the momenta. Considering their projection on the Minkowski space:
\begin{equation}
\label{c118}
(p,\,q)\rightarrow (p\oplus q)=(p_{a}\,e^{a}_{\;\mu}(p)+q_{a}\,\tilde{e}^{a}_{\;\mu}(q))
\end{equation}
and the generalization for the composition of a generic number of different species momenta:
\begin{equation}
\label{c119}
(p,\,q,\,k\ldots)\rightarrow (p\oplus q\oplus k\oplus\ldots)=(p_{a}\,e^{a}_{\;\mu}(p)+q_{a}\,e'^{a}_{\;\mu}(q)+k_{a}\,e''^{a}_{\;\mu}(k)+\ldots)
\end{equation}
From the fact that different particle species momenta live in different space-time, the momenta conservation must be valid only on the common support space-time, that is this rule must be preserved only for the momenta local projection on the Minkowski support space:
\begin{equation}
\label{c119a}
\begin{split}
&p_{(1)\mu}e^{\mu}_{\;j}(p_{(1)})+p_{(2)\mu}e^{\mu}_{\;j}(p_{(2)})=q_{(1)\mu}e^{\mu}_{\;j}(q_{(1)})+q_{(2)\mu}e^{\mu}_{\;j}(q_{(2)})\;\Rightarrow\\
\Rightarrow\;&p_{(1)j}+p_{(2)j}=q_{(1)j}+q_{(2)j}
\end{split}
\end{equation}
where $p_{(1)}$ and $p_{(2)}$ represent the incoming momenta and  $q_{(1)}$ and $q_{(2)}$ are the outgoing ones.\\
Therefore HMSR model, proposed in this work, presents an analogy with DSR theories \cite{AmelinoCamelia,AmelinoCamelia2}, but it is important to underline a difference. In fact the modified composition rule does not present an universal character, instead it is species depending, and moreover it is associative and abelian. The new physics emerges by the comparison of different particle species that have different Modified Lorentz Transformations. The construction of the modified physics can, therefore, predict physical effects, experimentally detectable. Instead in case of a physics modification with universal character, independent from the particle species, new physical effects correspond to a redefinition of the units of measure - that is the speed of light \cite{Liberati3}.

\section{Standard Model modifications}
As underlined in \cite{Torri,Antonelli}, the introduction of a deformed geometry, influences the form of the Dirac equation, with the result of modifying spinors and correlated currents. The deformed Dirac matrices can be computed, requiring that they satisfy the Clifford Algebra relation:
\begin{equation}
\label{c51}
\{\Gamma_{\mu},\Gamma_{\nu}\}=2\,g_{\mu\nu}(p)=2\,e_{\mu}^{\,a}(p)\,\eta_{ab}\,e_{\nu}^{\,b}(p)
\end{equation}
from which it is simple to obtain the equality:
\begin{equation}
\label{c52}
\Gamma^{\mu}=e^{\;\mu}_{a}(p)\,\gamma^{a}
\end{equation}
From the previous equation it is immediate to compute the modified Dirac matrices explicit forms:
\begin{equation}
\label{c53}
\begin{split}
&\Gamma_{0}=\gamma_{0}\qquad \Gamma_{i}=\frac{1}{\sqrt{1-f(p(x,\,\dot{x}))}}\;\gamma_{i}\simeq\sqrt{1+f(p(x,\,\dot{x}))}\;\gamma_{i}\\
&\Gamma^{0}=\gamma^{0}\qquad \Gamma^{i}=\sqrt{1-f(p(x,\,\dot{x}))}\;\gamma^{i}
\end{split}
\end{equation}
The $\Gamma_{5}$ matrix can be introduced using the total antisymmetric tensor $\epsilon_{\mu\nu\alpha\beta}$, defined in curved space-time:
\begin{equation}
\label{c54}
\begin{split}
&\Gamma_5=\frac{\epsilon^{\mu\nu\alpha\beta}}{4!}\Gamma_{\mu}\Gamma_{\nu}\Gamma_{\alpha}\Gamma_{\beta}=\frac{1}{\sqrt{\det{g}}}\Gamma_{0}\Gamma_{1}\Gamma_{2}\Gamma_{3}=\\
=&\frac{1}{\sqrt{\det{g}}}\sqrt{\det{g}}\,\gamma_{0}\gamma_{1}\gamma_{2}\gamma_{3}=\gamma_{5}
\end{split}
\end{equation}
As consequence the new constructed geometry preserves the standard chirality classifications of particles.\\
To determine the explicit form of spinors and associated conserved currents, first it is essential to define the modified Dirac equation:
\begin{equation}
\label{c55}
\left(i\Gamma^{\mu}\partial_{\mu}-m\right)\psi=0
\end{equation}
From this equation, following a standard argumentation, present in literature \cite{Schwabl}, it is possible to obtain the modified spinors. Assuming the possibility to develop the general spinor in plane waves:
\begin{equation}
\label{c56}
\begin{split}
&\psi^{+}(x)=u_{r}(p)e^{-ip_{\mu}x^{\mu}}\\
&\psi^{-}(x)=v_{r}(p)e^{ip_{\mu}x^{\mu}}
\end{split}
\end{equation}
and taking into account only the positive energy one (for the negative one the computation retains the same form), the modified spinors can be easily computed from the associated Dirac equation in momentum space. Applying this equation to the generic positive energy spinor, it is possible to obtain:
\begin{equation}
\label{c57}
(i\Gamma^{\mu}\partial_{\mu}-m)u_{r}(p)e^{-ip_{\mu}x^{\mu}}\Rightarrow(\slashed{p}-m)u_{r}(p)=0
\end{equation}
where the generalized relation $\slashed{p}=\Gamma^{\mu}p_{\mu}$ has been used. Finally it is simple to derive the associated identity for spinors with null momentum $\overrightarrow{p}=0$:
\begin{equation}
\label{c58}
\begin{split}
&(\slashed{p}-m)(\slashed{p}+m)=(p_{\mu}p_{\nu}g^{\mu\nu}(p)-m^2)=(p^{\mu} p_{\mu})-m^2=0\;\Rightarrow \\
\Rightarrow&\;(\slashed{p}-m)(\slashed{p}+m)u_{r}(m,\,\overrightarrow{0})=0
\end{split}
\end{equation}
From previous relation follows that generic momentum $\overrightarrow{p}$ spinors can be obtained from those with null momentum. From this statement, the possibility to compute modified positive energy not normalized spinor immediately follows. Starting from the null momentum positive energy spinor standard representation:
\begin{equation}
\label{c59}
u_{r}(m,\,\overrightarrow{0})=\chi_{r}=\left(
                                         \begin{array}{c}
                                           1 \\
                                           0 \\
                                         \end{array}
                                       \right)
\end{equation}
it is simple to compute the generic spinor from the relation:
\begin{equation}
\label{c60}
\begin{split}
&(\Gamma^{\mu}p_{\mu}+m)\left(
                          \begin{array}{c}
                            \chi_{r} \\
                            0 \\
                          \end{array}
                        \right)
\Rightarrow\\
\Rightarrow&
\left(p^{0}\left(
             \begin{array}{cc}
               \mathbb{I} & 0 \\
               0 & \mathbb{-I} \\
             \end{array}
           \right)-
p^{i}\left(
       \begin{array}{cc}
         0 & \sigma^{i} \\
         -\sigma^{i} & 0 \\
       \end{array}
     \right)
\sqrt{1-f}\right)
\left(
  \begin{array}{c}
    \chi_{r} \\
    0 \\
  \end{array}
\right)+\\
+&m\left(
   \begin{array}{cc}
     \mathbb{I} & 0 \\
     0 & \mathbb{I} \\
   \end{array}
 \right)
 \left(
   \begin{array}{c}
     \chi_{r} \\
     0 \\
   \end{array}
 \right)
 =\left(
     \begin{array}{c}
       (E+m)\chi_{r} \\
       \overrightarrow{p}\overrightarrow{\sigma}\sqrt{1-f}\;\chi_{r} \\
     \end{array}
   \right)
\end{split}
\end{equation}
and finally the modified spinor normalized form can be written as:
\begin{equation}
\begin{split}
\label{c61}
&\left(
     \begin{array}{c}
       (E+m)\chi_{r} \\
       \overrightarrow{p}\overrightarrow{\sigma}\sqrt{1-f}\;\chi_{r} \\
     \end{array}
\right)\;\Rightarrow\\
\Rightarrow&\;u_{r}(m,\,\overrightarrow{p})=
\frac{1}{\sqrt{2m(E+m)}}
\left(
     \begin{array}{c}
       (E+m)\,\chi_{r} \\
       \overrightarrow{p}\overrightarrow{\sigma}\,\sqrt{1-f}\,\chi_{r} \\
     \end{array}
\right)
\end{split}
\end{equation}
Having defined the modified spinors from the plane-waves expansion, it is now possible to verify its compatibility with the MDR, in fact:
\begin{equation}
\label{c62}
\begin{split}
&\left(i\Gamma^{\mu}\partial_{\mu}+m\right)\left(i\Gamma^{\nu}\partial_{\nu}-m\right)u(p)e^{-ip_{\mu}x^{\mu}}\Rightarrow\\
\Rightarrow&\left(\frac{1}{2}\{\Gamma^{\mu},\Gamma^{\nu}\}p_{\mu}p_{\nu}-m^2\right)u(p)=0\Rightarrow\\
\Rightarrow&\left(p_{\mu}p_{\nu}g^{\mu\nu}-m^2\right)u(p)=0\Rightarrow\\
\Rightarrow& E^2-|\overrightarrow{p}|^2(1-f(p))-m^2=0
\end{split}
\end{equation}
This proves that the free propagation of the introduced modified spinors is governed by the MDR (\ref{c4}).\\
To describe a physical theory like QED or SM (weak sector and QCD), it is essential to deal with interaction terms and therefore it is necessary to introduce the theory modified conserved currents. Starting from the simpler case of QED, the current must be defined as a spinor bilinear, in order to be contracted with the boson gauge vectorial field of the theory. Moreover spinorial bilinear and gauge boson field must be projected on the same tangent space, to permit this contraction. Therefore, assuming the gauge fields as Lorentz invariant, the introduced theory contemplates a kinematical modification, but not a dynamical one. That is the new aspects are limited to the kinematics of the free particles, without modifying the known interactions. This can be achieved by the introduction of the generalized $\widetilde{\Gamma}$ matrices:
\begin{equation}
\label{c63}
\widetilde{\Gamma}_{\mu}(p',\,p)=\left(
                                     \begin{array}{cc}
                                        0 & \sigma_{a}e^{a}_{\;\mu}(p') \\
                                        \overline{\sigma}_{a}e^{a}_{\;\mu}(p) & 0 \\
                                     \end{array}
                                 \right)
\end{equation}
and the consequent modified current is given by:
\begin{equation}
\label{c64}
J_{\mu}=e\sqrt{|\det{[\widetilde{g}]}|}\;\;\overline{\psi}\,\widetilde{\Gamma}_{\mu}(p,\,p')\,\psi
\end{equation}
where $e$ is the coupling constant (the electric charge), $p$ represents the incoming spinor field momentum and $p'$ the outgoing one, and the generalized metric has been introduced:
\begin{equation}
\label{c65}
\{\widetilde{\Gamma_{\mu}}(p,\,p'),\,\widetilde{\Gamma_{\nu}}(p,\,p')\}=2\,\widetilde{g}_{\mu\nu}(p,\,p')
\end{equation}
In this way, the LIV corrections present in the modified spinors compensate the modified matrix ones. The current lives in the tanget space $(TM,\,\eta_{\mu\nu})$ and therefore it is possible to write:
\begin{equation}
\label{c66}
J^{\mu}=\eta^{\mu\nu}\,J_{\nu}
\end{equation}
The interaction term can therefore be written, in the most general form, as:
\begin{equation}
\label{c67}
\mathcal{L}_{inter}=e\sqrt{|\det{[\widetilde{g}]}|}\;\;\overline{\psi}\,\widetilde{\Gamma}_{\mu}(p,\,p')\,\psi\,\overline{e}^{\mu}_{\;\nu}\,A^{\nu}
\end{equation}
where $\overline{e}$ represents the projector (\emph{virerbein} element) correlated to the gauge field and the index $\mu$, even if greek, represents a coordinate of the Minkowski space-time $(TM,\,\eta_{\mu\nu})$. Under the gauge fields Lorentz covariance assumption, the $\overline{e}$ thetrad reduces to the form $\overline{e}_{\mu}^{\;\nu}=\delta_{\mu}^{\;\nu}$ and the gauge bosons live in the flat space-time.  The term, that in (\ref{c64}) and (\ref{c67}) multiplies the conserved current, is a generalization of the analogous term borrowed from curved space-time QFT, where its explicit form is given by: $\sqrt{|\det{[g]}|}$ \cite{Parker}.
With the previous definitions it is possible now to write the interaction Lagrangians of the LIV perturbed theories, that become for the QED:
\begin{equation}
\label{c68}
\mathcal{L}=\sqrt{|\det{[g]}|}\;\;\overline{\psi}(i\Gamma^{\mu}\partial_{\mu}-m)\psi+e\sqrt{|\det{[\widetilde{g}]}|}\;\;\overline{\psi}\,\widetilde{\Gamma}_{\mu}(p,\,p')\,\psi\,\overline{e}^{\mu}_{\;\nu}\,A^{\nu}
\end{equation}
where $\widetilde{g}(p,\,p')$ is obtained in (\ref{c65}) and $g(p)$ represents the metric computed in (\ref{c14}) that coincides with that used in (\ref{c51}).\\
It is important to underline that the modified Dirac matrices (\ref{c52}) are used to write the kinetic part of the Lagrangian, describing the free fermion propagation. This part determines the form of the propagator of the particle and, as consequence, the dispersion relation, that is the MDR (\ref{c4}).\\
In the low energy scenario the perturbations result negligible. Instead, in case of high energy limit, it is possible to consider incoming and outgoing momenta with approximately the same magnitude, even after interaction. Therefore $\widetilde{\Gamma}$ matrices admit a constant form high energy limit and can be assumed not depending on the momenta. The definition of the current (\ref{c64}) reduces, as in \cite{Torri,Antonelli}, to:
\begin{equation}
\label{c69}
J_{\mu}=e\sqrt{|\det{\frac{1}{2}\{\Gamma_{\mu},\,\Gamma_{\nu}\}}|}\;\;\overline{\psi}\,\Gamma_{\mu}\,\psi=e\sqrt{|\det{[g]}|}\;\;\overline{\psi}\,\Gamma_{\mu}\,\psi
\end{equation}
because $\widetilde\Gamma_{\mu}(p,\,p')\rightarrow\Gamma_{\mu}$ if $p\simeq p'$ and therefore $\widetilde{g}_{\mu\nu}(p,\,p')\rightarrow g_{\mu\nu}(p)$. Since the perturbation magnitude is supposed tiny and its effects are visible only for high energies, the last formulation can be considered as the main one. Moreover it is a reasonable physical hypothesis to suppose the quantum effects, caused by the interaction with the background, tiny for massless particles and it is possible to neglect this contribution for the gauge field $A_{\mu}$. In this way the gauge field results Lorentz invariant and preserves even the gauge symmetry. The Lagrangian becomes:
\begin{equation}
\label{c70}
\mathcal{L}=\sqrt{\det{[g]}}\;\overline{\psi}(i\Gamma^{\mu}D_{\mu}-m)\psi
\end{equation}
with the theory local covariant derivative defined as:
\begin{equation}
\label{c70a}
D_{\mu}=\partial_{\mu}-ie\,A_{\mu}
\end{equation}
Important to stress that:
\begin{equation}
\label{c70b}
\sqrt{\det{[g]}}\;\overline{\psi}\,\Gamma^{\mu}\,\psi=\sqrt{\det{[g]}}\;\overline{\psi}\,\Gamma_{\nu}\,\psi\,\eta^{\mu\nu}
\end{equation}
thanks to (\ref{c66}), because the current is defined in a Minkowskian space-time. With the $\widetilde{\Gamma}$ defined as in (\ref{c63}), the matrices corrections are compensated by the spinor fields ones. Using the $\Gamma$ matrices, defined in (\ref{c52}), corrections survive, caused by the difference of incoming and outgoing momenta, but these perturbations are negligible. Therefore spinor corrections cancel the gamma matrices ones, or only negligible corrections survive, permitting to assume that the current itself lives in a flat space-time.\\
The same generalization can be applied to the SM Lagrangian, using again the vierbein correlated to different fermions to project them on the common support Minkowski space-time. Even if one is not dealing with mass eigenstates, it is possible to consider the perturbation again as function of momentum and energy ratio (\ref{c4}).\\
Using the chirality projectors, it is possible to define, in the usual way, the left and right-hand component for every particle field:
\begin{equation}
\label{c71}
\begin{split}
&\psi_{L}=P_{L}\,\psi=\frac{1}{2}\left(\mathbb{I}-\Gamma_{5}\right)\,\psi=\frac{1}{2}\left(\mathbb{I}-\gamma_{5}\right)\,\psi\\
&\psi_{R}=P_{R}\,\psi=\frac{1}{2}\left(\mathbb{I}+\Gamma_{5}\right)\,\psi=\frac{1}{2}\left(\mathbb{I}+\gamma_{5}\right)\,\psi
\end{split}
\end{equation}
where the equality $\gamma_{5}=\Gamma_{5}$ has been used.\\
The left-handed neutrino-lepton $(\nu-l)$ flavor $f$ doublets can opportunely be defined in the usual fashion, as:
\begin{equation}
\label{c72}
L^{f}_{L}=      \left(
                  \begin{array}{c}
                    \nu^{f}_{L} \\
                    l^{f}_{L} \\
                  \end{array}
                \right)=\left(
                \left(
                  \begin{array}{c}
                    \nu^{e}_{L} \\
                    e_{L} \\
                  \end{array}
                \right),\;
                \left(
                  \begin{array}{c}
                    \nu^{\mu}_{L} \\
                    \mu_{L} \\
                  \end{array}
                \right),\;
                \left(
                  \begin{array}{c}
                    \nu^{\tau}_{L} \\
                    \tau_{L} \\
                  \end{array}
                \right)\right)
\end{equation}
the right-handed leptons:
\begin{equation}
\label{c73}
R^{f}=(l^{f})_{R}=\left(e_{R},\,\mu_{R},\,\tau_{R}\right)
\end{equation}
Analogously one can introduce the left-handed quark up-down $(u-d)$ flavor $f$ doublets as:
\begin{equation}
\label{c74}
Q^{f}_{L}=\left(
                  \begin{array}{c}
                    u^{f}_{L} \\
                    d^{f}_{L} \\
                  \end{array}
                \right)=\left(
                \left(
                  \begin{array}{c}
                    u_{L} \\
                    d_{L} \\
                  \end{array}
                \right),\;
                \left(
                  \begin{array}{c}
                    c_{L} \\
                    s_{L} \\
                  \end{array}
                \right),\;
                \left(
                  \begin{array}{c}
                   t_{L} \\
                   b_{L} \\
                  \end{array}
                \right)\right)
\end{equation}
and the right-handed up-down $(u-d)$ quark as:
\begin{equation}
\label{c75}
\left((u^{f})_{R},\;(d^{f})_{R}\right)
\end{equation}
Starting from the leptonic part, the weak interaction Lagrangian can be written, for the free propagation part as:
\begin{equation}
\label{c76}
\begin{split}
\mathcal{L}_{free}=&\,\sqrt{|\det{[g(L^{f})]}|}\left(i\,\overline{L}^{f}\,\Gamma(L^{f})^{\mu}\,\partial_{\mu}\,L^{f}\right)+\\
+&\,\sqrt{|\det{[g(R^{f})]}|}\left(i\,\overline{R}^{f}\,\Gamma(R^{f})^{\mu}\,\partial_{\mu}\,R^{f}\right)
\end{split}
\end{equation}
where the modified $\Gamma$ matrices (\ref{c52}) have been used and are defined by the particle species of the interaction considered. Therefore they depends on the left-handed doublet or on the right-handed lepton flavor.\\
The neutral current interaction term can be written as:
\begin{equation}
\label{c77}
\begin{split}
\mathcal{L}_{n.c.}=&\,\sqrt{|\det[{\widetilde{g}(\psi^{f})}]|}\;\bigg(i\,\overline{\psi}^{f}\,g_{0}\,\frac{Y}{2}\,\widetilde{\Gamma}(\psi)^{\mu}\,\psi^{f}\,\widetilde{e}(B)_{\mu}^{\,\nu}\,B_{\nu}+\\
+&\,i\,\overline{\psi}^{f}\,g_{1}\,\frac{\tau^{0}}{2}\,\widetilde{\Gamma}(\psi)^{\mu}\,\psi^{f}\,\widetilde{e}(W^{0})_{\mu}^{\,\nu}\,W^{0}_{\nu}\bigg)
\end{split}
\end{equation}
where $\psi$ represents every leptonic field, both left and right-handed, $g_{0}$ and $g_{1}$ are the coupling constants and $Y$ and $\tau^{0}$ are the usual matrices, in diagonal form, correlated respectively with the $U(1)$ and the $SU(2)$ gauge symmetries. $\widetilde{e}(B)_{\mu}^{\nu}$ and $\widetilde{e}(W^{0})_{\mu}^{\nu}$ are the vierbein correlated respectively with the gauge fields $B_{\mu}$ and $W^{0}_{\mu}$. As for QED, in order to guarantee that the neutral currents live in the tangent space $(TM,\,\eta_{\mu\nu})$, the interaction terms must be written using the generalized $\widetilde{\Gamma}$ matrices (\ref{c63}), that depend on the particle $\psi$.\\
The Lagrangian charged current interaction term instead acquires the explicit form:
\begin{equation}
\label{c78}
\begin{split}
&\mathcal{L}_{c.c.}=\\
&=g_{1}\,\sqrt{|\det{[\widetilde{g}(L^{f})]}|}\left(i\,\overline{L}^{f}\,\widetilde{\Gamma}(L^{f})^{\mu}\,\tau^{+}\,L^{f}\,\widetilde{e}(W^{+})_{\mu}^{\,\nu}\,W^{+}_{\,\nu}\right)+\\
&+h.c.
\end{split}
\end{equation}
where the matrices $\tau^{+}=\frac{1}{2}(\tau^{1}+i\tau^{2})$ and $\tau^{-}=\frac{1}{2}(\tau^{1}-i\tau^{2})$ are correlated respectively to the gauge fields $W^{+}=\frac{1}{2}(W^{1}-iW^{2})$ and $W^{-}=\frac{1}{2}(W^{1}+iW^{2})$ and are again in diagonal form. The $\widetilde{\Gamma}$ matrices (\ref{c63}), depending from the field $L^{f}$, have been used to define the interaction terms, for the same reason illustrated before.\\
As for the QED case, considering the high energy limit, when interaction term incoming and outgoing momenta are approximately the same, the generalized $\widetilde{\Gamma}$ matrices (\ref{c63}) reduce to the modified $\Gamma$ (\ref{c52}). Moreover, assuming also in this case the coupling of the gauge fields with the background negligible, the interaction preserves the gauge symmetry. Therefore it is possible to write the Standard Model leptonic Lagrangian as:
\begin{equation}
\label{c79}
\mathcal{L}_{lept}=\sqrt{|\det{[g^{(f)}]}|}\left(\overline{L}^{f}\,i\,\Gamma^{\mu}\,D_{\mu}\,L^{f}+\overline{R}^{f}\,i\,\Gamma^{\mu}\,D_{\mu}\,R^{f}\right)
\end{equation}
introducing the $SU(2)\times U(1)$ covariant derivative $D_{\mu}$:
\begin{equation}
\label{c79a}
D_{\mu}=\partial_{\mu}-ig_{0}\frac{Y}{2}B_{\mu}-ig_{1}\frac{\tau^{i}}{2}W^{i}_{\mu}
\end{equation}
and posing the $\Gamma$ matrices and the metric $g$ dependents only on the particle flavor and not on the particle chirality.\\
The quark sector weak interaction Lagrangian can be written in an similar fashion, with the free propagation term given by:
\begin{equation}
\label{c80}
\begin{split}
\mathcal{L}_{free}=&\sqrt{\det{[g(Q_{L}^{f})]}}\left(i\,\overline{Q}_{L}^{f}\,\Gamma(Q_{L}^{f})^{\mu}\,\partial_{\mu}\,Q_{L}^{f}\right)+\\
+&\,\sqrt{\det{[g(u_{R}^{f})]}}\left(i\,\overline{u}_{R}^{f}\,\Gamma(u_{R}^{f})^{\mu}\,\partial_{\mu}\,u_{R}^{f}\right)+\\
+&\,\sqrt{\det{[g(d_{R}^{f})]}}\left(i\,\overline{d}_{R}^{f}\,\Gamma(d_{R}^{f})^{\mu}\,\partial_{\mu}\,d_{R}^{f}\right)=\\
=&\,\sqrt{|\det{[g^{(f)}]}|}\,\Bigl(\overline{Q}^{f}\,i\,\Gamma_{(f)}^{\mu}\,\partial_{\mu}\,Q^{f}+\\
+&\,\overline{u}_{R}^{f}\,i\Gamma_{(f)}^{\mu}\,\partial_{\mu}\,u_{R}^{f}+\overline{d}^{f}_{R}\,i\,\Gamma_{(f)}^{\mu}\,\partial_{\mu}\,d^{f}_{R}\Bigr)
\end{split}
\end{equation}
with the $\Gamma$ matrices, and consequently the metric, depending only on the quark flavor and being equal for the same doublet left-handed quarks.\\
The neutral current interaction term can be written again as in (\ref{c77}) with the fields $\psi$ that represent left handed doublets and right handed singlet quarks fields.
To write the charged current term one must take into account that this interaction is not diagonal in the chosen quark fields basis. The explicit form of this term becomes:
\begin{equation}
\label{c81}
\begin{split}
&\mathcal{L}_{c.c.}=\\
&=g_{1}\sqrt{|\det{[\widetilde{g}_{(fg)}]}|}\,\left(i\,\overline{Q}^{f}\,\widetilde{\Gamma}^{\mu}_{(fg)}\,\tau^{i}_{fg}\,Q^{g}\,\widetilde{e}(W^{i})_{\mu}^{\,\nu}\,W^{i}_{\,\nu}\right)+\\
&+h.c.
\end{split}
\end{equation}
where $\tau^{i}_{fg}$ are the interaction matrices correlated to the gauge fields $W^{i}$ with $i=\pm$. The $\widetilde{\Gamma}$ matrices have been generalized to take into account the not diagonal coupling of quark doublets, and are defined as:
\begin{equation}
\label{c82}
\widetilde{\Gamma}^{\mu}_{(fg)}=\left(
                                  \begin{array}{cc}
                                    0 & \sigma^{a}\,e_{a(f)}^{\,\mu}(p) \\
                                    \overline{\sigma}^{a}\,e_{a(g)}^{\,\mu}(p') & 0 \\
                                  \end{array}
                                \right)
\end{equation}
where the vierbein $e_{a(f)}^{\,\mu}(p)$ is correlated to a quark doublet of flavor $f$, with momentum $p$. The generalized metric, generated by these modified matrices, takes the form:
\begin{equation}
\label{c83}
\{\widetilde{\Gamma}_{(f)}^{\mu}(p),\,\widetilde{\Gamma}_{(g)}^{\nu}(p')\}=2\,\widetilde{g}^{\mu\nu}_{(fg)}(p,\,p')
\end{equation}
This metric again defines the space-time where the interaction takes place, that is where the conserved current propagates and the interaction vertex is defined.\\
Even for quark sector, the high energy limit can be treated considering that the $\widetilde{\Gamma}_{(fg)}$ matrices tend to a constant form, not depending on the momenta. They maintain only the dependence on the doublet flavor correlated. This permits to write this interaction term as:
\begin{equation}
\label{c84}
\mathcal{L}_{c.c.}=g_{1}\sqrt{|\det{[g_{(fg)}]}|}\,\left(i\,\overline{Q}^{f}\,\Gamma^{\mu}_{(fg)}\,\tau^{i}_{fg}\,Q^{g}\,W^{i}_{\mu}\right)
\end{equation}
again supposing the perturbation effects, correlated to the gauge fields, negligible.\\
It is interesting to compare the introduced modified $\Gamma$ matrices with the form:
\begin{equation}
\label{c85}
\Gamma^{\mu}_{(fg)}=\eta^{\mu\nu}+c^{\mu\nu}_{(fg)}\gamma_{\nu}
\end{equation}
as in \cite{Koste2}, to show that the high energy limit corresponds to a redefinition of the metric, as in the cited work, where the modified metric is defined as $g_{\mu\nu}=\eta_{\mu\nu}+c_{\mu\nu}$.\\
Even the quark sector weak interaction Lagrangian can be written using the $SU(2)\times U(1)$ gauge covariant derivative $D_{\mu}$ (\ref{c79a}), obtaining an explicit form, similar to eq. (\ref{c79}). Following the same methodology used till now, it is possible to modify even the strong interaction Lagrangian, obtaining for the interaction term:
\begin{equation}
\label{c86}
\begin{split}
\mathcal{L}&_{strong}=\\
&=g_{s}\sqrt{|\det{[\widetilde{g}(p,\,p')]}|}\,\left(i\,\overline{Q}^{(f)}_{i}\,\widetilde{\Gamma}_{f}^{\;\mu}(p,\,p')\,t^{a}_{ij}\,Q^{(f)}_{j}\,\overline{e}(G)_{\mu}^{\,\nu}G_{\nu}^{a}\right)
\end{split}
\end{equation}
with $t^{a}$ indicating the matrix form of the generators of $SU(3)$ gauge symmetry group, with $i$ and $j$ representing the colour indices of the quark fields, $g_{s}$ the strong coupling constant and $\overline{e}(G)_{\mu}^{\,\nu}$ represents the projector (tetrad) correlated with the $G_{\nu}^{a}$ gauge field. Even in this case the Lagrangian can be rewritten in the simpler form:
\begin{equation}
\label{c87}
\mathcal{L}_{strong}=g_{s}\sqrt{|\det{[g]}|}\,\left(i\,\overline{Q}^{(f)}_{i}\,\Gamma_{f}^{\;\mu}\,t^{a}_{ij}\,Q^{(f)}_{j}\,G_{\mu}^{a}\right)
\end{equation}
using again the constant high energy limit of the $\widetilde{\Gamma}$ matrices and the $\widetilde{g}$ metric.\\
Posing again the gauge fields (in this case the gluons) Lorentz covariant, the complete formulation of the amended Standard Model can be simplified, introducing the $SU(3)\times SU(2)\times U(1)$ covariant derivative $D_{\mu}$:
\begin{equation}
\label{c89a}
D_{\mu}=\partial_{\mu}-ig_{0}\frac{Y}{2}B_{\mu}-ig_{1}\frac{\tau^{i}}{2}W^{i}_{\mu}-ig_{s}t^{i}G^{i}_{\mu}
\end{equation}
and resorting to the modified Dirac matrices, preserving the gauge formulation of the theory.\\
The last Lagrangian part to be amended remains the gauge free propagation fields terms. This part can be modified in a similar way as done in \cite{Koste2} and can be written as:
\begin{equation}
\label{c88}
\begin{split}
\mathcal{L}_{gauge}=&\,\frac{1}{4}\,g^{(G)}_{\mu\nu}\,g^{(G)}_{\alpha\beta}\,Tr(G^{\mu\alpha}G^{\nu\beta})\,+\\
+&\,\frac{1}{4}\,g^{(W)}_{\mu\nu}\,g^{(W)}_{\alpha\beta}\,Tr(W^{\mu\alpha}W^{\nu\beta})\,+\\
+&\,\frac{1}{4}\,g^{(ph)}_{\mu\nu}\,g^{(ph)}_{\alpha\beta}\,B^{\mu\alpha}B_{\nu\beta}
\end{split}
\end{equation}
where the metric $g^{f}_{\mu\nu}$ depends on the gauge field $f$ species considered, and $\{G_{\mu\nu}$, $W_{\mu\nu}$, $B_{\mu\nu}\}$  represent the gauge fields strength. The similarity with \cite{Koste2} is given by the tensor that appears in the perturbation term $k^{(f)}_{\mu\nu\alpha\beta}=g^{(f)}_{\mu\nu}\,g^{(f)}_{\alpha\beta}-\eta_{\mu\nu}\eta_{\alpha\beta}$. Supposing the gauge field interaction with the background negligible, this Lagrangian term reduces to the standard form:
\begin{equation}
\label{c89}
\mathcal{L}_{gauge}=\frac{1}{4}\,Tr(G^{\mu\nu}G_{\mu\nu})+\frac{1}{4}\,Tr(W^{\mu\nu}W_{\mu\nu})+\frac{1}{4}B^{\mu\nu}B_{\mu\nu}
\end{equation}
Finally it is not necessary to introduce perturbations in Higgs sector and in the Yukawa interaction term, to provide a kinematical modification that affects only fermions.

\section{Allowed symmetries}
The SM modifications, introduced in HMSR, are conceived in order to preserve space-time homogeneity and isotropy, but even the standard physics interactions and. As consequence, the same $SU(3)\times SU(2)\times U(1)$ internal symmetries are preserved. To prove this statement it is possible to verify that the Coleman-Mandula theorem \cite{Coleman2} is still valid. In this way it results that the allowed symmetries are restricted to the direct product of internal ones with those generated by the modified Lorentz group, introduced before. A less rigorous proof can be obtained generalizing a Witten argument \cite{Argyres} about the fact that any additional kinematic and non internal symmetry would overconstrain the scattering amplitude. Therefore any further symmetry generator beyond Lorentz group would allow nontrivial scattering amplitude only for a discrete set of scattering angles.\\
It is possible to start from admitting the existence of a symmetry generator $Q_{\mu\nu}$, symmetric, traceless and such that:
\begin{equation}
\label{c90}
[Q_{\mu\nu},\,P_{\alpha}]\neq 0\;\;and\;\;Q_{\mu\nu}\neq J_{\mu\nu}\in\mathfrak{so}(1,\,3)
\end{equation}
$\forall P_{\alpha}$ generator of the Poincar\'e group.\\
The symmetry and tracelessness of $Q_{\mu\nu}$ let to write:
\begin{equation}
\label{c91}
\langle p|Q_{\mu\nu}|p\rangle\div p_{\mu}p_{\nu}-\frac{1}{4}g_{\mu\nu}(p)\,p^2
\end{equation}
where the tracelessness is evaluated using the metric $g_{\mu\nu}(p)$ (\ref{c22}). Moreover, assuming that this operator acts like a tensor, for orthonormal states $|p_{(1)}\rangle$ and $|p_{(2)}\rangle$, one obtains the equality:
\begin{equation}
\label{c92}
\langle p_{(1)},\,p_{(2)}\,|Q_{\mu\nu}|\,p_{(1)},\,p_{(2)}\rangle=\langle p_{(1)}\,|Q_{\mu\nu}|\,p_{(1)}\rangle+\langle p_{(2)}\,|Q_{\mu\nu}|\,p_{(2)}\rangle
\end{equation}
Using the momentum conservation defined in (\ref{c119a}), if the outgoing momenta projected on the support space are: $q_{(1)j}=p_{(1)j}+a_{j}$ and $q_{(2)j}=p_{(2)j}+b_{j}$, from the momentum conservation for elastic scattering $p_{(1)j}+p_{(2)j}=q_{(1)j}+q_{(2)j}$, one can obtain $a_{j}=-b_{j}$. From the $Q_{\mu\nu}$ conservation it follows now that:
\begin{equation}
\label{c93}
\langle p_{(1)},\,p_{(2)}\,|Q_{\mu\nu}|\,p_{(1)},\,p_{(2)}\rangle=\langle q_{(1)},\,q_{(2)}\,|Q_{\mu\nu}|\,q_{(1)},\,q_{(2)}\rangle
\end{equation}
and from this relation, using (\ref{c91}) and making a series expansion for the metric $g_{\mu\nu}(p)$, using the tetrad elements as projectors, it is possible to write the relation:
\begin{equation}
\label{c94}
\begin{split}
&p_{(1)\mu}e^{j}_{\;\nu}(p_{(1)})a_{j}+p_{(1)\nu}e^{j}_{\;\mu}(p_{(1)})a_{j}+\\
+&\,e^{j}_{\;\mu}(p_{(1)})a_{j}e^{k}_{\;\nu}(p_{(1)})a_{k}-p_{(2)\mu}\widetilde{e}^{j}_{\;\nu}(p_{(2)})a_{j}+\\
-&\,p_{(2)\nu}\widetilde{e}^{j}_{\;\mu}(p_{(2)})a_{j}+\widetilde{e}^{j}_{\;\mu}(p_{(2)})a_{j}e^{k}_{\;\nu}(p_{(2)})a_{k}+\\
-&\,\frac{1}{2}\frac{\partial}{\partial p_{(1)\alpha}}g_{\mu\nu}(p_{(1)})p_{(1)}^{\;2}e^{j}_{\;\alpha}(p_{(1)})a_{j}+\\
-&\,\frac{1}{2}\frac{\partial}{\partial p_{(2)\alpha}}g_{\mu\nu}(p_{(2)})p_{(2)}^{\;2}\widetilde{e}^{j}_{\;\alpha}(p_{(2)})a_{j}=0
\end{split}
\end{equation}
Since the derivative $\partial_{\alpha}g_{\mu\nu}(p)$ are negligible and the vierbeins $e^{j}_{\;\mu}(p_{(1)})\sim\mathbb{I}$ and $\widetilde{e}^{j}_{\;\mu}(p_{(2)})\sim\mathbb{I}$, from (\ref{c94}) equation it follows that $a_{j}=0\;\Rightarrow\;a_{\mu}=0$ and this means that only trivial scattering is allowed.

\section{Coleman-Mandula theorem generalization}
Following the demonstration present in \cite{Weinberg}, it is possible to verify that the theorem is still valid, even replacing the underlying Minkowski geometry with the pseudo-Finsler, considered in this work, and generalizing the Lorentz group, which acquires a dependence on the particle momentum. Therefore the theorem hypotheses must be modified respect to those present in \cite{Weinberg} and can be written as:
\begin{enumerate}
  \item Lorentz invariance respect to the Modified Lorentz Transformations
  \item Particle number finitness: $\forall M>0$ $\exists n<\infty$ number of particles with mass $m<M$
  \item Elastic scattering is an analytic function of the modified Mandelstam variables
  \item Nontrivial scattering happens for almost all energies
  \item $\forall g\in G$, where $G$ is the symmetry group, the element $g\in U(1)$ is representable in a identity neighbourhood via an integral operator, with distribution kernel
\end{enumerate}
The $S$ matrix is expressed as a function of the modified Mandelstam variables and is therefore invariant under the action of the modified Lorentz group.\\
The first part of the demonstration regards the subset of symmetry operators, that commute with the Poincar\'e group. Generalizing the classic idea of operators that act as tensors on different particle states, one obtains:
\begin{equation}
\label{c94a}
B_{\alpha}=\otimes_{i}B_{\alpha}^{(i)}
\end{equation}
where the $i$ index is related to the different particle species taken into account. These operators therefore satisfy the relation:
\begin{equation}
\label{c95}
[B_{\alpha}^{\,(i)},\,P_{\mu}^{\,(i)}]=0\quad\forall B_{\alpha}^{\,(i)}\in G_{sym},\;\forall P_{\mu}^{\,(i)}\in\mathcal{P}
\end{equation}
where $G_{sym}$ is the symmetry group. These operators act therefore on single particle states as:
\begin{equation}
\label{c96}
\begin{split}
B_{\alpha}&|p,\,m,\,q,\,n\ldots\rangle=\otimes_{i}B_{\alpha}^{\,(i)}|p,\,m,\,q,\,n\ldots\rangle=\\
=&\sum_{m'}[b_{\alpha}^{\,i}(p)]_{mm'}|p,\,m',\,q,\,n\ldots\rangle+\\
+&\sum_{n'}[b_{\alpha}^{\,i}(p)]_{nn'}|p,\,m,\,q,\,n'\ldots\rangle+\ldots
\end{split}
\end{equation}
where $[b_{\alpha}^{\,i}(p)]_{mm'}$ is the matrix representation of the operator $B_{\alpha}$, relative to the $i$ particle species. Since the operators commute with the Poincar\'e group generators, as in the classical case, they satisfy the Lie algebra commutation rules:
\begin{equation}
\label{c97}
\begin{split}
&[B_{\alpha}^{\,(i)},\,B_{\beta}^{\,(i)}]=i\,C_{\alpha\beta}^{\tau(i)}\,B_{\tau}^{\,(i)}\\
&[b_{\alpha}^{\,(i)}(p),\,b_{\beta}^{\,(i)}(p)]=i\,C_{\alpha\beta}^{\tau(i)}\,b_{\tau}(p)^{\,(i)}\quad\forall p
\end{split}
\end{equation}
Now starting from this relation it is possible to follow the classic demonstration, to prove that the correspondence $B_{\alpha}^{\,(i)}\longrightarrow[b_{\alpha}^{\,i}(p)]$ is a bijection and therefore even $B_{\alpha}\longrightarrow\otimes_{i}[b_{\alpha}^{\,i}(p)]$. It is only necessary to be careful to replace the on-shell condition of a particle with the MDR (\ref{c4}) and considering that:
\begin{equation}
\label{c98}
\begin{split}
&\langle p',\,m',\,q',\,n'|[B_{\alpha},\,S]| p,\,m,\,q,\,n\rangle=\\
&=\langle p',\,m',\,q',\,n'|[\otimes_{i}B_{\alpha}^{\,(i)},\,S]| p,\,m,\,q,\,n\rangle=0
\end{split}
\end{equation}
since the operators $B_{\alpha}$ are symmetry generators and commute with the $S$ matrix, function of the new defined Mandelstam variables.
Moreover the computation of the particle number with mass lower than a given number is:
\begin{equation}
\label{c99}
N_{i}\left(\sqrt{p_{\mu}p^{\mu}}\right)=N_{i}\left(\sqrt{p_{\mu}\,g_{i}^{\,\mu\nu}(p)\,p_{\nu}}\right)
\end{equation}
so the particle number finitness is still preserved. Again the $i$ index represents the particle species, taken into account. Now, as in the classical version of the theorem, it is possible to find operators:
\begin{equation}
\label{c100}
B^{\sharp(i)}_{\alpha}=B_{\alpha}^{\,(i)}-a_{\alpha}^{\mu(i)}P_{\mu}^{\,(i)}
\end{equation}
for opportune coefficients $a_{\alpha}^{\mu(i)}$. These operators commute with $P_{\mu}^{\,(i)}$ and $[P_{\mu}^{\,(i)},\,J^{(i)}(p)]$, where $J^{(i)}(p)$ is a generator of the modified Lorentz group, for given momentum and given $i$ particle species. The last statement is true because $[P_{\mu}^{\,(i)},\,J^{(i)}(p)]$ is given by a linear combination of $P_{\mu}^{\,(i)}$ momenta, so the Jacobi identity:
\begin{equation}
\label{c101}
\begin{split}
[P_{\mu}^{\,(i)},&\,[J^{(i)}(p),\,B^{\sharp(i)}]]+[J^{(i)}(p),\,[B^{\sharp(i)},\,P_{\mu}^{\,(i)}]]+\\
+&[B^{\sharp(i)},\,[P_{\mu}^{\,(i)},\,J^{(i)}(p)]]=0
\end{split}
\end{equation}
is still valid. Now it is possible to show that:
\begin{equation}
\label{c102}
\begin{split}
&[B^{\sharp(i)}_{\alpha},J^{(i)}(P)]=0\;\Rightarrow\\\
\Rightarrow\;&[\otimes_{i}B^{\sharp(i)}_{\alpha},\otimes_{i}J^{(i)}(P)]=0\;\Rightarrow\\
\Rightarrow\;&[B^{\sharp}_{\alpha},J(P)]=0
\end{split}
\end{equation}
proving the theorem for the case of operators belonging to this particular subalgebra.\\
Considering now the symmetry generators subgroup, made of operators that do not commute with the Poincar\'e group: $[A_{\alpha},\,P_{\beta}]\neq0$, one can write the action of a generic element of this group, on a single particle state, as:
\begin{equation}
\label{c103}
A^{(i)}_{\;\alpha}|p_{(i)},\,n_{(i)}\rangle=\sum_{n'}\int\,d^{4}p'\left[\mathcal{A}^{(i)}_{\;\alpha}(p,\,p')\right]_{nn'}|p'_{(i)},\,n'_{(i)}\rangle
\end{equation}
where again the index $i$ represents the particle species.\\
The classical theorem demonstration version focuses on the fact that this kind of operators have integral kernel null for $p\neq p'$. This remains valid even in the modified case, considering again the modified version of the mass-shell definition. The argumentation remains the same, arriving to demonstrate that such an operator can be written as:
\begin{equation}
\label{c104}
A^{(i)}_{\;\alpha}=-\frac{i}{2}a(p)_{\alpha}^{(i)\mu\nu}J^{(i)}_{\;\mu\nu}+B^{(i)}_{\;\alpha}
\end{equation}
with an opportune coefficient $a(p)_{\alpha}^{(i)\mu\nu}$, proving that this type of symmetry generators are given by the direct product of Poincar\'e group elements times internal symmetry generators:
\begin{equation}
\label{c105}
\begin{split}
&A^{(i)}=P^{(i)}(p)\otimes G_{sym}\;\Rightarrow\\
\Rightarrow\;&A=\otimes_{i}A^{(i)}=\otimes_{i}P^{(i)}(p)\otimes G_{sym}
\end{split}
\end{equation}
Finally it is possible to state that the allowed symmetries of the scenario, proposed in HMSR, are given by the direct product $\mathcal{P}(p)\otimes G_{int}$, where $\mathcal{P}(p)$ is the direct product of modified Poincar\'e groups, that depends explicitly on the particle species and energy (momentum):
\begin{equation}
\label{c105a}
\mathcal{P}(p)=\otimes_{i}\mathcal{P}^{(i)}(p_{(i)})
\end{equation}
and $G_{int}$ is the internal symmetries group (in this case $SU(3)\times SU(2)\times U(1)$).

\section{SME correspondence}
The introduction of species depending MLT permits to introduce new Physics, generated by the different way particles are affected by LIV.  As already underlined in \cite{Koste5}, this idea is compatible with SME, where different particles can break the Lorentz symmetry differently. LIV is introduced in HMSR, starting from a kinematical modification, that can be investigated by the isotropic coefficients of the SME. In this work only MDRs that are equal for particles and antiparticles have been considered. This corresponds to modify the Standard Model introducing only CPT-even therms, as illustrated in \cite{Koste2}. Furthermore the MDRs form selected does not distinguish between particle polarizations. In fact, considering a SM extension with CPT even terms of the form:
\begin{equation}
\label{c120}
\frac{1}{2}i\,c_{\mu\nu}\overline{\psi}\gamma^{\mu}\overleftrightarrow{D}^{\nu}\psi+\frac{1}{2}i\,d_{\mu\nu}\overline{\psi}\gamma_{5}\gamma^{\mu}\overleftrightarrow{D}^{\nu}\psi
\end{equation}
it is possible to define the modified Dirac matrices:
\begin{equation}
\label{c121}
\Gamma^{\mu}=\gamma^{\mu}+c^{\mu\nu}\gamma_{\nu}+d^{\mu\nu}\gamma_{5}\gamma_{\nu}
\end{equation}
obtaining an effective Lagrangian that induces an MDR with a difference, taking into account that one is dealing with real fermions (particles with spin). The present work considers a subset of SME, the one generated by the isotropic coefficient $c_{\mu\nu}$. Moreover it introduces isometry transformations for this subclass of violation cases, in order to preserve space-time isotropy. The only difference with the SME theory consists in posing the trace of this coefficient not null: $Tr\,(c_{\mu\nu})\neq0$. This hypothesis is not considered in SME, because it represents
a simple scaling of the kinetic term and therefore is only part of the definition of the normalization of
the field. In other words the trace of this tensor represents a universal modification of the maximum attainable velocity, eventuality that in SME is supposed to not generate visible physical effects. Instead in the present work it is proved that the species depending character of the MLT can generate detectable effects.

\section{Conclusions}
A possible way to introduce a standard model extension, that preserves the idea of isotropy, is the key idea of HMSR and this work. As already highlighted, this is possible taking into account some concepts of the SME \cite{Koste2} with some ideas borrowed from DSR \cite{AmelinoCamelia,AmelinoCamelia2}. The key point consists in constructing the space-time starting from a pure kinematical modification, that results depending on the particle species. This leads to a new space-time structure, that depends on the propagating material body momentum, the Finsler geometry. The Lorentz invariance is not broken, but modified, introducing an amended Lorentz group, in order to redefine the concept of spatial symmetries and reconcile the introduced perturbations of space-time with the idea of symmetry conservation. Moreover the perturbation considered have only a kinetic character, so the dynamic is not affected and new exotic interactions are not introduced, preserving the internal $SU(3)\times SU(2)\times U(1)$ standard model symmetry. In this way it is simple to generalize the concept of isotropy, respect to the new generalized personal Lorentz transformations. The physical effects of such a theory can emerge only in interaction processes where different particle species are involved. In fact a universal modification would generate, for example, a redefinition of the measure units, effect very difficult to be detected. Instead, as shown in \cite{Torri,Antonelli}, processes like GZK cut-off and neutrino oscillations, where different particle species interact, can be affected by this LIV model.

\section{Aknowledgements}
A special thanks to prof. Alan Kostelecky for reading the draft of this work, suggesting some relevant improvements, and thanks to all the partecipants of the Third IUCSS Summer School and Workshop on Lorentz- and CPT-violating Standard-Model Extension, for the interesting and useful physics conversations.

\end{document}